\documentclass[%
 reprint,
 amsmath,amssymb,
 aps,
]{revtex4-1}

\usepackage{dsfont}
\usepackage{graphicx}
\usepackage{dcolumn}
\usepackage{bm}
\usepackage{hyperref}
\usepackage{subfig}
\usepackage{microtype}
\usepackage{tikz}
\usepackage{pgfplots, pgfplotstable}
\usepackage{tkz-graph}
\usetikzlibrary{arrows}
\usetikzlibrary{matrix}
\usetikzlibrary{patterns}
\usetikzlibrary{shapes}
\usetikzlibrary{shapes.multipart}
\usepgfplotslibrary{groupplots}
\usetikzlibrary{pgfplots.groupplots}

\tikzstyle{na} = [baseline=-.5ex]
\tikzstyle{nb} = [baseline=-1.2ex]

\definecolor{orange}{rgb}{1,0.3,0.1}
\definecolor{lila}{rgb}{0.9,0,0.3}
\definecolor{green}{rgb}{0,0.6,0}
\definecolor{hell}{rgb}{0.6,0.6,0}
\definecolor{white}{rgb}{1.0,1.0,1.0}

\renewcommand{\vec}{\boldsymbol}

\newcommand{\vecS}[3]{{\stackrel{_\Delta}{#1}}_{#2}^{#3}}

\newtheorem{remark}{Remark}
\newtheorem{defi}{Definition}
\newtheorem{example}{Example}

\graphicspath{{./Fig/}}

\begin{document}

\preprint{APS/123-QED}

\title{Continuous Time Limits of the Utterance Selection Model}

\author{J\'er\^ome Michaud}
 \email{v1jmicha@staffmail.ed.ac.uk}
\affiliation{%
 School of Physics, University of Edinburgh, Mayfield Road, Edinburgh EH9 3JZ, United Kingdom
}%




\date{\today}

\begin{abstract}
In this paper, we derive new continuous time limits of the Utterance Selection Model (USM)  for language change (Baxter et al., Phys. Rev. E {\bf 73}, 046118, 2006). This is motivated by the fact that the Fokker-Planck continuous time limit derived in the original version of the USM is only valid for a small range of parameters. We investigate the consequences of relaxing these constraints on parameters. Using the normal approximation of the multinomial approximation, we derive a new continuous time limit of the USM in the form of a weak-noise stochastic differential equation. We argue that this weak noise, not captured by the Kramers-Moyal expansion, can not be neglected. We then propose a coarse-graining procedure, which takes the form of a stochastic version of the \emph{heterogeneous mean field} approximation. This approximation groups the behaviour of nodes of same degree, reducing the complexity of the problem. With the help of this approximation, we study in detail two simple families of networks: the regular networks and the star-shaped networks. The analysis reveals and quantifies a finite size effect of the dynamics. If we increase the size of the network by keeping all the other parameters constant, we transition from a state where conventions emerge to a state when no convention emerges. Furthermore, we show that the degree of a node acts as a time scale. For heterogeneous networks such as star-shaped networks, the time scale difference can become very large leading to a noisier behaviour of highly connected nodes.
\end{abstract}

\pacs{Valid PACS appear here}
\maketitle

\section{Introduction}
In the study of complex systems, one important challenge is to deduce the macroscopic behaviour of a system from the microscopic dynamics. This problem is at the centre of statistical mechanics. In this paper, we are interested in the (stochastic) agent-based class of complex systems. In (stochastic) agent-based models, agents interact following some rules (subject to noise) and we would like to characterize the averaged behaviour of the complete population. One possibility to obtain a characterization of the averaged behaviour is by obtaining a mean field approximation. What is meant by a mean field approximation varies between authors. The original idea is to  characterize the dynamics of a complex system by choosing a representative agent and approximating the effect of the rest of the population as a mean field, see for example \cite{kadanoff2009more}. This approach is well-adapted to well-mixed populations, but in the case of heterogeneous populations, for example when the social structure is a complex network, this approach usually fails to describe the dynamics. To tackle this problem, the heterogeneous mean field approximation (HMF) has been proposed. In this approximation, the dynamics of agents in a network is approximated by taking one representative agent for each degree class. For more details on the HMF approximations and other approximations of the dynamics on complex network, the reader is referred to \cite{gleeson2013binary}. For some application of the HMF for different models, the interested reader is referred to \cite{sood2005voter,sood2008voter,Castellano2005,newman2002spread}. These two mean field approaches are based on the averaged influence of the different group considered and provide a deterministic approximation of the dynamics. They share the property to average out the detail of the underlying structure of the interactions.

In this paper, we present an alternative to the usual mean field approaches by keeping some stochasticity in the HMF approximation. In the HMF approximation, one uses degree-block variables to estimate the dynamics. This is only one of many possible choices to introduce an heterogeneity in the mean field approach. Alternatively, one can group the agents by community or by any relevant criteria instead of by degree. An HMF approximation can then be obtained by using block variables, where the blocks depend on the grouping criteria. This procedure does not imply a deterministic approximation and some stochasticity can be conserved in the coarse-grained approximation and we will refer to this novel approximation as the stochastic HMF (sHMF).

As an example, we apply the sHMF procedure to the problem of language evolution.
Language is a defining property of humanity and is at the centre of human interactions. The study of language dynamics is very important to better understand the formation of human cultures. In particular, the dynamics of language contacts and the formation of new dialects, pidgins or creoles can shed light on the mechanisms underlying the formation and evolution of socio-cultural groups, see for example \cite{tria2015modeling}. Language is a complex adaptive system \cite{BeBlByChCrElHoKeLaSch09,Steels00} and can be described at many different scales \cite{Michaud16, ChristiansenKirby03}. It seems that the different scales of language evolution should be accounted for in a better way than it has previously been done. In fact, at the interaction scale languages are highly variable, whereas at the population scale languages are relatively stable and change on a slow time scale. In order to better understand the link between these two time scales a coarsening procedure such as the new sHMF approximation is needed.


Here, we focus on the specific instance of the Utterance Selection Model (USM) for language change \cite{BaxterEtAl06} and
derive an sHMF approximation of it.
The USM is a stochastic agent-based model describing the evolution of a population interacting by stochastically producing utterances and learning from them. Although there exists a wide range of models of language evolution, see \cite{Grifoni15}, we find the USM particularly appealing in
 that it can describe the process of language language both at the timescale of individual interactions and at the timescale of the population. The USM has been applied to evaluate the theory of Trudgill of the emergence of New Zealand English \cite{BaxterEtAl09}.  Under appropriate assumptions, this model is analytically tractable and a wide range of results regarding are available. The main results on the dynamics of the USM have been obtained in 
 \cite{BaxterEtAl06,BlytheCroft12,PopFrey13} and we review them below.
 
 In \cite{BaxterEtAl06}, continuous time limits at the interaction level have been obtained using Kramers-Moyal (KM) expansion and provide an analytical tool to study the marginal distribution of a representative agent in a population. However, in order to obtain this continuous time limit, one has to restrict the parameter space to simplify the mathematics. In order to fully characterize the behaviour of the model, this restriction on parameters has to be overcome.

In \cite{BlytheCroft12}, modifications of the USM are investigated in order to characterize under which circumstances language change trajectories follow a so-called S-curve. Linguistic corpora studies \cite{pemberton1936curve,GhanbarnejadEtAl2014} have shown that language change trajectories typically follow S-curves.

 Finally, in \cite{PopFrey13} scaling law for the time needed to achieve consensus are obtained and numerically validated. This paper is one of the few considering parameter values outside the range in which the results of \cite{BaxterEtAl06} are valid. In this paper, we extend and improve previously known results by obtaining a novel continuous time limit of the USM at the interaction time scale, which does not suffer from any parameter restriction. We also obtain a coarse-grained sHMF approximation of the USM, clarifying the conditions under which a consensus can be achieved in this model.

The remaining of this paper is organized as follows. In Sec.~\ref{sec:TimeScale} we discuss the coarse-graining problematic and clarify our strategy to obtain a sHMF of the USM. In Sec.~\ref{sec:USM} we recall the definition of the USM and some known results. In Sec.~\ref{sec:SDE} we derive a weak-noise stochastic differential equation (SDE) generalizing the continuous time limit obtained in \cite{BaxterEtAl06} and compare the numerical efficiency of different numerical algorithms. This shows that for two agents, the system mainly behaves in a deterministic manner for short times, but stochastic effect become relevant in long time scales. In Sec.~\ref{sec:HMF} we derive the sHMF approximation of the USM and apply it to regular and star-shaped networks to validate it. This allows us to obtained a mean field characterization of the noise-driven phase transition separating the conditions under which a consensus can or cannot be formed at the population level. The analysis reveals a finite size effect justifying the fact that in small population it is easier to create conventions. In Sec.~\ref{sec:Conclusion} we summarize the main results of this research and discuss the future research directions. This paper is complemented by four appendices. In App.~\ref{App:abb} the abbreviations used in this paper are collected. In App.~\ref{App:USM}, we rederive the continuous time limit obtained in \cite{BaxterEtAl06} for completeness. In App.~\ref{App:WF} the USM is linked with Wright-Fisher process and technical details about the SDE are provided. Finally, in App.~\ref{App:NumSim} details on numerical methods used to numerically integrate the sHMF equations are provided.

\section{\label{sec:TimeScale}Time scale separation and coarse-graining problem}
\begin{figure}[h]
\centering
\includegraphics{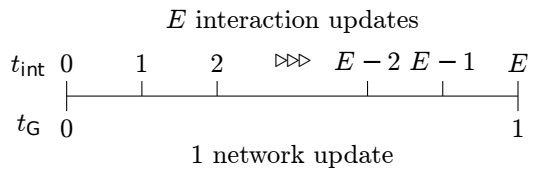}
\caption{Illustration of the two timescales of the problem. $t_{\rm int}$ represents the time of one interaction and $t_{\mathcal G}$ represents the time of $1$ network update.}
\label{fig:timescales}
\end{figure}

In this section, we discuss the time scale separation problem inherent to every agent-based model and set out the approach taken in this paper. In order to simplify the discussion, we consider the case in which agents are associated with vertices $\mathcal V$ of a static network $\mathcal G$. Assuming pairwise interactions, such a system possesses two natural time scales: an interaction time scale $t_{\rm int}$ and a network time scale $t_{\mathcal G}$. Imagine that a clock, associated with $t_{\rm int}$, ticks at every new interaction  (assuming sequential updates) and that another clock, associated with $t_{\mathcal G}$, ticks when all the edges of the graph have been updated, then, on average, the interaction clocks ticks $E$ times between two ticks of the network clock, where $E$ is the number of edges of the graph. This situation is illustrated in Fig.~\ref{fig:timescales}.

If the number of edges $E$ is large, the time scale separation between $t_{\rm int}$ and $t_{\mathcal G}$ increases. In fact, the relationship between these two time scales is
\begin{equation}
t_{\rm int} \approx \frac1{E}t_{\mathcal G}.
\end{equation}
 In the limit when $E\to \infty$ the dynamics at the interaction level can be considered as continuous, since $t_{\rm int}\to 0$.  This motivates the need to develop continuous time limits of the dynamics at the agent level in order to derive a population level approximation of the dynamics.
If the network is finite, we expect some finite size effects to occur, modifying the dynamics.

As we have mentioned, we aim to obtain a coarse-grained approximation of the dynamics of an agent-based model (ABM) and we would this approximation to be continuous in the network time $t_{\mathcal G}$ for large enough network. There are therefore two problems that need to be solved: the coarse-graining problem and the continuous time limit problem.  


\begin{figure}[h]
\centering
\includegraphics{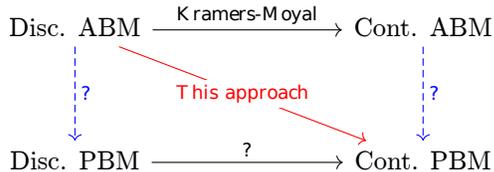}
\caption{Illustration of the coarsening problem by a continuous time approximation. The discrete models are on the left. Agent-based models are on top, their evolution depends on $t_{\rm int}$ and on the bottom the population models evolve according to $t_{\mathcal G}$. For the USM, it is known how to obtain a continuous time limit at the agent level using the Kramers-Moyal expansion. The other arrows are not clear. In this paper we will take the diagonal approach.}
\label{fig:CoarseContinuous}
\end{figure}

In Fig.~\ref{fig:CoarseContinuous} we provide an illustration of this problem. We want to derive a continuous in time ($t_{\mathcal G}$) population-based model (PBM) starting from a discrete in time agent-based model (ABM). One can first coarsen the problem and then obtain a continuous time limit or do the opposite. In this paper we will do both in a single step. As mentioned in Fig.~\ref{fig:CoarseContinuous}, for the USM the only approximation that has been studied is a continuous in time approximation at the agent level in the form of a KM expansion leading to a Fokker-Planck (FP) equation, see \cite{BaxterEtAl06} and App.~\ref{App:USM} for details. This approximation suffers from parameter restrictions and cannot be easily coarse-grained. It is not clear how one can in general approximate the other arrows for the USM. In this paper, we both provide an alternative to the KM expansion, obtaining a continuous time limit at the agent level without parameters restrictions, and a methodology to derive a continuous in time population-based approximation in the form of a sHMF.




\section{\label{sec:USM}The Utterance Selection Model}
We now recall the definition of the USM.
The USM \cite{BaxterEtAl06} is a stochastic agent-based model of language evolution based on an evolutionary theory of language change due to Croft \cite{Croft00}. This model is not limited to the cultural evolution of languages but can be interpreted as a general model of cultural evolution. 

In this USM, $N$ agents are represented as nodes of a static network $\mathcal G =(\mathcal V, \mathcal E)$, where $\mathcal V$ is the set of vertices and $\mathcal E$ the set of edges along which the agents interact. We assume this network to be undirected and weigthed by a probability distribution $G^{(ij)}$ representing the probability that agent $i\in \mathcal V$ interacts with agent $j\in \mathcal V$. In order to model the cultural evolution of a trait, the USM assumes that a particular trait can be instantiated in $V$ equivalent variants. The state of an agent is characterized by a probability distribution $\vec x$ over the possible $V$ variants of the cultural trait, which can be interpreted as her belief of  the frequency with which she should use the variants. 
 In other words, $\vec x$ models her \emph{idiolect} and cannot be accessed by other agents.
 Since $\vec x$ is a discrete probability distribution it belongs to
\begin{equation}\label{eq:PV}
\mathds P_V:= \left\{\vec x \in [0,1]^V | \sum_{v=1}^{V} x_v = 1\right\}.
\end{equation}
  In order to communicate, an agent produces an utterance $\vec u\in\mathds P_V$ from a production process $\mathcal U$ ($\vec u:=\mathcal U \vec x$), which takes the form of an  empirical distribution of a biased sample of length $L$ of her belief distribution or idiolect. The length of the utterance $L$ controls the amount of variability in the speech, since when $L$ is large, the utterances are long and the induced noise small. The biasing process models production errors and/or innovation and is encoded through a stochastic matrix $M$. The updating (or learning) rule is formed by the weighted average  of a process of self-monitoring $S$ (weighted by $(1-h^{(ij)})$) and a process of accommodation $A$ (weighted by $h^{(ij)}$). The process of self-monitoring aims at reducing the difference between $\vec x^{(i)}$ and $\vec u^{(i)}$ of an agent $i$ and the accommodation process aims at reducing the difference between $\vec x^{(i)}$ and the utterance $\vec u^{(j)}$ of an neighbouring agent $j$. The model is completed by a small parameter $\lambda$ modelling the rate of learning. The interpretation of the parameters of the USM is summarized in Tab.~\ref{tab:USMparam}.

\begin{table}[h]
\caption{Interpretation of the parameters of the USM}
\label{tab:USMparam}
\begin{tabular}{cl}
\hline\hline
Parameter&Interpretation\\
\hline
$N$ & Number of agents\\
$V$&Number of variants\\
$M$&Innovation\\
$L$&Variability\\
$G^{(ij)}$ & Probability of interaction\\
$h^{(ij)}$&Attention parameter\\
$\lambda$&Learning rate\\
\hline\hline
\end{tabular}
\end{table}

An interaction time step of the USM can be divided into three substeps: social interaction, utterance production and retention. A simulation run of the USM iterates such an interaction time step $ET$ times, where $E$ is the number of edges of the network and $T$ is the final time of the simulation in $t_{\mathcal G}$ units. The three substeps of an interaction time step are defined below.
\begin{description}
\item[Social interaction] The social interaction is simply modelled by choosing a pair of speakers $i,j$ with the prescribed probability $G^{(ij)}$. In this paper, we only consider the case where $G^{(ij)} = \frac1{E}$, that is, the uniform distribution. Furthermore, in order to be closer to the discussion about time scales of Sec.~\ref{sec:TimeScale}, instead of randomly sampling the edges, we randomly order them and go through them in sequence in such a way that when a network update is complete, all the edges have been updated exactly once.
\item[Utterance production] The production phase is illustrated in Fig.~\ref{fig:DUSM} by the $\mathcal U$ operator.
\begin{figure}[h]
\includegraphics{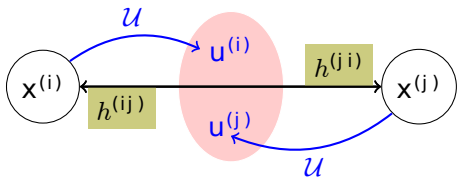}
\caption{Structure of the USM interaction. On the $(ij)$ edge, the agents use their internal beliefs $\vec x^{(i)}$ to produce an utterance $\vec u^{(i)}$ through the process $\mathcal U$, which depend on the matrix $M$. The utterances are then used to update the internal beliefs depending on a weighting parameter $h^{(ij)}$.}
\label{fig:DUSM}
\end{figure}
 It occurs at a specified time $t_{\rm int}$. The two chosen agents generate an utterance $\vec u^{(i)}$. 
 The sampling process is done by using a multinomial sampling and the biasing process is done through the introduction of a mutation matrix $M$, which is column stochastic.
 Note that the ordering of the sampling and the biasing processes matters. We therefore have the two possible definitions of the utterance empirical frequency vector $\vec u^{(i)}$:
\begin{subequations}
\label{eq:utter}
\begin{eqnarray}\label{eq:utterBS}
 \vec u_{\rm bs}&\sim&\frac{1}{L}\text{Multi}(L, M\vec x);\label{eq:utterBS}\\
 \vec u_{\rm sb}&\sim&\frac{1}{L}M\text{Multi}(L, \vec x).\label{eq:utterSB}
\end{eqnarray}
\end{subequations}

In \cite{BaxterEtAl06}, the rule \eqref{eq:utterBS} has been chosen to model the utterance process. We argue in App.~\ref{App:WF} that the other choice \eqref{eq:utterSB} is more natural and leads to a well-posed SDE, whereas the choice \eqref{eq:utterBS} leads to an ill-posed SDE. In \cite{BaxterEtAl06}, the differences between this two choices are lost during the derivation of the continuous time limit. If the specific rule is not specified, we use the notation $\mathcal U \vec x = \vec u$ without subscript.

The different utterances produced during a communication event form an utterance pool on which the retention phase is based.
\item[Retention] The retention rule, or updating rule, is a rule to compute $\vec x^{(i)}(t+1)$, where $t$ is measured in $t_{\rm int}$ units. This is the short timescale updating rule. 
An agent $i$ then revises her state $\vec x^{(i)}$ using
\begin{equation}\label{eq:DUSM}
\begin{aligned}
\delta \vec x^{(i)}(t)= \lambda\Big[&\left(1-h^{(ij)}\right)S\left(\vec x^{(i)}(t),\vec u^{(i)}(t)\right)\\& + h^{(ij)}A\left(\vec x^{(i)}(t),\vec u^{(j)}(t)\right)\Big],
\end{aligned}
\end{equation}
where $\delta \vec x^{(i)}(t) = \vec x^{(i)}(t+1)-\vec x^{(i)}(t)$. 
In this description, the utterance vectors $\vec u$ are stochastic vectors.
We define the self-monitoring process $S$ and the accommodation process $A$ as 
\begin{equation}\label{eq:SA}
\left\{
\begin{array}{lcl}
S\left(\vec x^{(i)}(t),\vec u^{(i)}(t)\right) &:=&\vec u^{(i)}(t)-\vec x^{(i)}(t),\\
A\left(\vec x^{(i)}(t),\vec u^{(j)}(t)\right) &:=&\vec u^{(j)}(t)-\vec x^{(i)}(t).
\end{array}
\right.
\end{equation}

These two processes are driven by probability matching, since they compare the empirical frequencies $\vec u$ with their belief probability distribution $\vec x$ until they match. The term \emph{probability matching} is widely used by evolutionary linguists to express situations in which speakers adapt their speech distribution to the speech distribution they hear. If the probability distributions are equal, then those terms vanish. Some variants of the USM use a different definition for the self-monitoring and accommodation processes. In \cite{BlytheCroft12}, the influence of misperception is investigated. This is outside the scope of this paper, but extending our results to these more general cases is in principle possible. 
 In this paper, we restrict the discussion to the original choice of probability matching.

 The relative weight of these two functions is given by a parameter $h^{(ij)}\in [0,1]$ and $\lambda>0$ is a usually small positive parameter. For simplicity, we assume that $h^{(ij)} = h$, that is, the attention parameter does not depend on the identity of agents. 
\end{description}

The complete mathematical definition of the discrete USM  is then given by Eqs~\eqref{eq:utter}--\eqref{eq:SA}. The USM contains two sources of randomness: the first is contained in the distribution $G^{(ij)}$, which controls the way in which the edges are updated; the second is contained in the utterance process $\mathcal U$, which controls the noisy interaction between agents. In order to characterize the model, one is interested in the statistical behaviour, which can be studied through approximations. Continuous time limits deal with the noisy utterance process, whilst coarse-graining approximations deal with the social noise. 

In \cite{BaxterEtAl06}, a FP equation has been obtained as an agent-level continuous time limit of the USM using KM expansion. In App.~\ref{App:USM} we recall this procedure and show that the required scaling assumptions (Eq.~\eqref{eq:scalingUSM}) significantly restrict the application of this approximation. In the next section, we derive an alternative continuous time limit of the USM based on a normal approximation of the multinomial distribution (diffusion approximation), which does not suffer from any parameter restriction and which generalizes the result obtained with the KM expansion.



\section{\label{sec:SDE}SDE continuous time limits}
In this section, we develop the first main contribution of this paper, that is, we derive a new continuous time limit of the USM that captures the dynamics of the USM over the full range of parameters. The limit is derived at the interaction time scale $t_{\rm int}$ and generalizes the FP equation obtained by KM expansion.



In the rest of this section, we first obtain approximations of the utterance production process. We then derive the weak-noise SDE continuous time limit of the USM. 
Finally, we test the different approximations against the discrete USM and against the deterministic limit obtained by the KM expansion with scaling $\lambda \propto \delta t$ on a very simple network and argue that the weak-noise should not be neglected.

\subsection{\label{sec:approx}Approximations of the multinomial distribution}
The USM utterance production mechanism given in Eq.~\eqref{eq:utter} relies on a multinomial sampling and a biasing procedure. In order to obtain a weak-noise SDE continuous time limit of the USM, we need (i) to approximate the sampling process in a continuous in $L$ manner and (ii) to decouple the parameters and the source of noise to relate the noise to a Wiener process. To do so, assume that we want to approximate a random vector 
\begin{equation}
\vec z \sim \frac1{L}{\rm Multi}(L,\vec y),
\end{equation}
where $L$ is an integer and $\vec y$ is a discrete probability vector,
by a vector $\vec w$. First note that the expectation value and covariance matrix of $\vec z$ are given by
\begin{subequations}
\begin{eqnarray}
\mathds E(\vec z) &=& \vec y,\\
{\rm Cov}(\vec z,\vec z) &=& \frac1{L} ({\rm diag}(\vec y)-\vec y \vec y^T).
\end{eqnarray}
\end{subequations}
A possible continuous in $L$ analog to the multinomial distribution is given by the Dirichlet distribution of parameter $L\vec y$ and we can approximate $\vec z$ by
\begin{equation}
\vec w^{\rm Dir}(\vec y) \sim {\rm Dir}(L\vec y).
\end{equation}
This approximation is continuous in $L$ but does not decouple the parameter and the noise source. The good property of this approximation is that $\vec w$ is a discrete probability distribution.

In order to decouple the source of noise from the parameter $\vec y$, one can use the normal approximation of the multinomial distribution. This leads to an approximation
\begin{equation}
\begin{aligned}\label{eq:Normalw}
\vec w^{\rm N}(\vec y) &\sim \mathds E(\vec z) + \left({\rm Cov}(\vec z,\vec z)\right)^{1/2} \mathcal N(0,\vec I)\\
&\sim \vec y + \frac{1}{\sqrt{L}}D(\vec y) \mathcal N(\vec 0,\vec I)
\end{aligned}
\end{equation}
where the square root has to be taken in the Cholesky sense and the matrix $D(\vec y)$ is the square root in the Cholesky sense of ${\rm diag}(\vec y) - \vec y \vec y^T$. A definition of a square root in the Cholesky sense is given in Def.~\ref{def:SCholesky} and the possible forms of the matrix $D(\vec y)$ are given in App.~\ref{App:WF}.

The normal approximation given by Eq.~\eqref{eq:Normalw} is both continuous in $L$ and decouples the source of noise and the parameter $\vec y$. This permits a connection with Wiener processes as will be shown below. 

The drawback of the normal approximation is that $\vec w^{\rm N}$ is not a discrete probability distribution vector in general. This is a consequence of the fact that the normal approximation is unbounded, whereas the multinomial and the Dirichlet distribution are bounded. We also have to note that this approximation is only valid if $L$ is sufficiently large and $\vec y$ not close to the boundaries of the domain. These assumptions are not always satisfied, but we will assume them anyway.

With these limitations in mind, one can now provide a continuous approximation of the utterance production process and introduce the biasing process through the matrix $M$. For the Dirichlet approximation and for the normal approximation, one can approximate the continuous in $L$ utterance vector as
\begin{subequations}\label{eq:ContLutter}
\begin{eqnarray}
\vec u^{\rm Type}_{\rm bs} &=&\vec w^{\rm Type}(M\vec x),\quad {\rm Type} \in \{{\rm Dir}, {\rm N}\}\label{eq:ContLutter.bs} \\
\vec u^{\rm Type}_{\rm sb} &=&M\vec w^{\rm Type}(\vec x),\quad {\rm Type} \in \{{\rm Dir}, {\rm N}\}
\end{eqnarray}
\end{subequations}
where $\vec x$ is the belief distribution state vector. The index bs stands for first biasing, then sampling and the index sb for the reverse ordering.

We will show in Sec.~\ref{ssec:wSDE} that under the normal approximation the order of application of the sampling and the biasing processes matters. This is connected to the fact that the continuous in $L$ utterance vector obtained under this approximation does not always represent a discrete probability distribution. 

For the normal approximation, the continuous in $L$ utterance vector are given by
\begin{subequations}\label{eq:NormalApp}
\begin{eqnarray}
\vec u^{\rm N}_{\rm bs} &=&M\vec x + \frac{1}{\sqrt{L}}D(M\vec x)\vec \xi, \\
\vec u^{\rm N}_{\rm sb} &=&M\vec x + \frac{1}{\sqrt{L}}MD(\vec x)\vec \xi,
\end{eqnarray}
\end{subequations}
where $\vec \xi \sim \mathcal N(\vec 0, \vec I)$.
\begin{remark}\label{rem:D}
It is mentioned in App.~\ref{App:USM} that one usually assumes that the off-diagonal terms of $M$ are small (of order $O((\delta t)^{1/2})$ or smaller). In that case, one can show in general that 
\begin{subequations}
\begin{eqnarray}
D(M\vec x) &= D(\vec x) + O(\|M-I\|_{\infty}),\\
MD(\vec x) &=D(\vec x) + O(\|M-I\|_{\infty}).
\end{eqnarray}
\end{subequations}
As a consequence, in the derivation of a continuous time equation, the influence of the matrix $M$ in the noise term can be neglected and the ordering between sampling and biasing no longer matters.
Note that using the  Dirichlet approximation produces a vector $\vec u^{\rm Dir}$ representing a discrete probability distribution under both orderings. This is also true for the discrete USM.
\end{remark}

\subsection{Weak-noise SDE limit}\label{ssec:wSDE}
We  have now collected all the partial results needed to derive continuous time limits of the USM and in particular a weak-noise SDE continuous time limit based upon the normal approximation. 

The derivation of the continuous time limits is now fairly staightforward, all that needs to be done is to put the continuous in $L$ approximations of the utterance vector into Eq. \eqref{eq:DUSM}, average over the possible interactions of an agent (average over its neighbours) and scale $\lambda = \delta t$ to obtain a continuous time limit.

For the Dirichlet approximation such an approximation is obtained by introducing the random vector $\vec u^{\rm Dir}$ defined in Eq.~\eqref{eq:ContLutter.bs}, with either the biasing-sampling order or the reverse order, into \eqref{eq:DUSM}, sum the contribution of all the neighbours of an agent $i$ and introducing the scaling $\lambda = dt$. We obtain
\begin{equation}\label{eq:DirContLim}
\begin{aligned}
\dot{\vec x}^{(i)}= \sum_{j\neq i}G^{(ij)}\Big[&\left(1-h\right)\left(\vec u^{{\rm Dir}(i)}-\vec x^{(i)}\right) \\ &+h\left(\vec u^{{\rm Dir}(j)}-\vec x^{(i)}\right)\Big].
\end{aligned}
\end{equation}
This is the first continuous time equation we consider. This is an SDE in the sense that the vectors $\vec u^{{\rm Dir}(i)}$ and $\vec u^{{\rm Dir}(j)}$ are stochastic vectors, but it is not a usual SDE, since the noise is not related to a Wiener process and cannot be analyzed in the framework of SDEs. We use this formulation in the numerical experiments as an accurate continuous time limit of the USM, since the random vector produced always represents a discrete probability distribution.


The derivation of the continuous time limit based on the normal approximation is obtained in a similar way as the Dirichlet approximation. We introduce the normal approximation \eqref{eq:NormalApp} into Eq.~\eqref{eq:DUSM}, sum the contribution of all the neighbours of an agent $i$ and introduce the scaling $\lambda = dt$.  Letting $dt \to 0$, we obtain the following two equations depending on the ordering choice between the biasing and the sampling processes.
\begin{subequations}
\label{eq:CTLSDE}
\begin{widetext}
\begin{equation}\label{eq:CTLSDE1}
d\vec x^{(i)}=\sum_{j\neq i}G^{(ij)} \Bigg[
\left((1-h)(M-I)\vec x^{(i)} +h(M\vec x^{(j)} -\vec x^{(i)})\right)dt + \left(\frac{1-h}{\sqrt{L}}D(M\vec x^{(i)})d\vec \xi_t^{(i)} + \frac{h}{\sqrt{L}}D(M\vec x^{(j)})d\vec \xi_t^{(j)}\right)\Bigg],
\end{equation}
or
\begin{equation}\label{eq:USMCTL}
d\vec x^{(i)}=\sum_{j\neq i}G^{(ij)} \Bigg[
\left((1-h)(M-I)\vec x^{(i)} +h(M\vec x^{(j)} -\vec x^{(i)})\right)dt + \left(\frac{1-h}{\sqrt{L}}MD(\vec x^{(i)})d\vec \xi_t^{(i)} + \frac{h}{\sqrt{L}}MD(\vec x^{(j)})d\vec \xi_t^{(j)}\right)\Bigg],
\end{equation}
\end{widetext}
\end{subequations}
where $d\vec\xi_t = \sqrt{dt}d\vec W_t \sim \mathcal N(\vec 0, dt^2\vec I)$. This noise is weaker than a usual gaussian noise $d\vec W_t$ by a factor $\sqrt{dt}$. We call this limit a weak-noise SDE. This approximation is a diffusion approximation taking into account all the sources of noise in an interaction, that is, the noise originating from the two utterances produced. This is different from the continuous time limit obtained in \cite{BaxterEtAl06}, where only the noise of the speaker is taken into account. Note that a deterministic limit is obtained by neglecting the noise terms in \eqref{eq:CTLSDE1} in which case the solution of the KM expansion when $\lambda = \delta t$ is recovered. This approximation, therefore, generalizes the KM expansion.

The coefficient of the noise scales as $\sqrt{dt}$ and vanishes in the continuous time limit in agreement with the FP derivation. We argue that the noise term of Eq.~\eqref{eq:CTLSDE} should not be neglected for two reasons. Firstly, since the white noise $dW_t$ scales as $\sqrt{dt}$, the noise term scales as $dt$ and can be argued to be of the same order of magnitude than the drift term. Secondly, the drift term has the property to become very small for long time, because $\vec x^{(i)}$ and $\vec x^{(j)}$ converge towards each other and they jointly converge toward the a vector $\vec b$, such that $(M-I)\vec b = 0$. As a result, even a weak noise becomes important in the long time as soon as the drift term becomes of order $\sqrt{dt}$. Therefore, we expect the noise to be important on long time scale, but negligible on short time scale. This will be verified with numerical simulations.


We argue that Eq.~\eqref{eq:CTLSDE1} is ill-posed and that Eq.~\eqref{eq:USMCTL} is well-posed. We recall the a problem is said to be well-posed if it has a unique solution and if small changes in initial conditions leads to small changes of the solution (stability). From Eqs~\eqref{eq:CTLSDE1} and \eqref{eq:USMCTL} it is not straightforward to decide whether there are well-posed or not. A detailed discussion of a special case of these equations is treated in App.~\ref{App:WF} and explain the origin of the ill-posedness of Eq.~\eqref{eq:CTLSDE1}. In the following, we will work with the continuous time limit Eq.~\eqref{eq:USMCTL}.


If we consider the scaling \eqref{eq:scalingUSM} instead of scaling only $\lambda$, the corresponding SDE reads
\begin{equation}\label{eq:SDEUSM}
\begin{aligned}
d\vec x^{(i)} = \sum_{j\neq i}G^{(ij)}\Big[&\left((\bar M-I)\vec x^{(i)} + \bar h(\vec x^{(j)}-\vec x^{(i)})\right)dt\\
&+\frac{1}{\sqrt{L}}D(\vec x^{(i)})d\vec W_t^{(i)}\Big],
\end{aligned}
\end{equation}
where $\vec W_t^{(i)}$ is a standard vectorial Wiener process and $d\vec W_t^{(i)}$ is a white noise. Eq. \eqref{eq:SDEUSM} is the stochastic counterpart of the FP equation obtained in \cite{BaxterEtAl06} using the It\={o} convention. The deterministic limit corresponds to scaling only $\lambda$, which is consistent with the FP equation derived by the KM expansion in \cite{BaxterEtAl06}. In this limit, the noise of agent $j$ becomes irrelevant and can be neglected. 
\begin{remark}
Eq. \eqref{eq:SDEUSM} is the same for the two possible orderings of the production, thanks to Rem. \ref{rem:D}. In other words, with the scaling used in \cite{BaxterEtAl06}, the two orderings become equivalent.
\end{remark}

\subsection{Numerical experiments}
\begin{figure*}[ht]
\centering
  \subfloat
  {\includegraphics[width=0.25\textwidth]{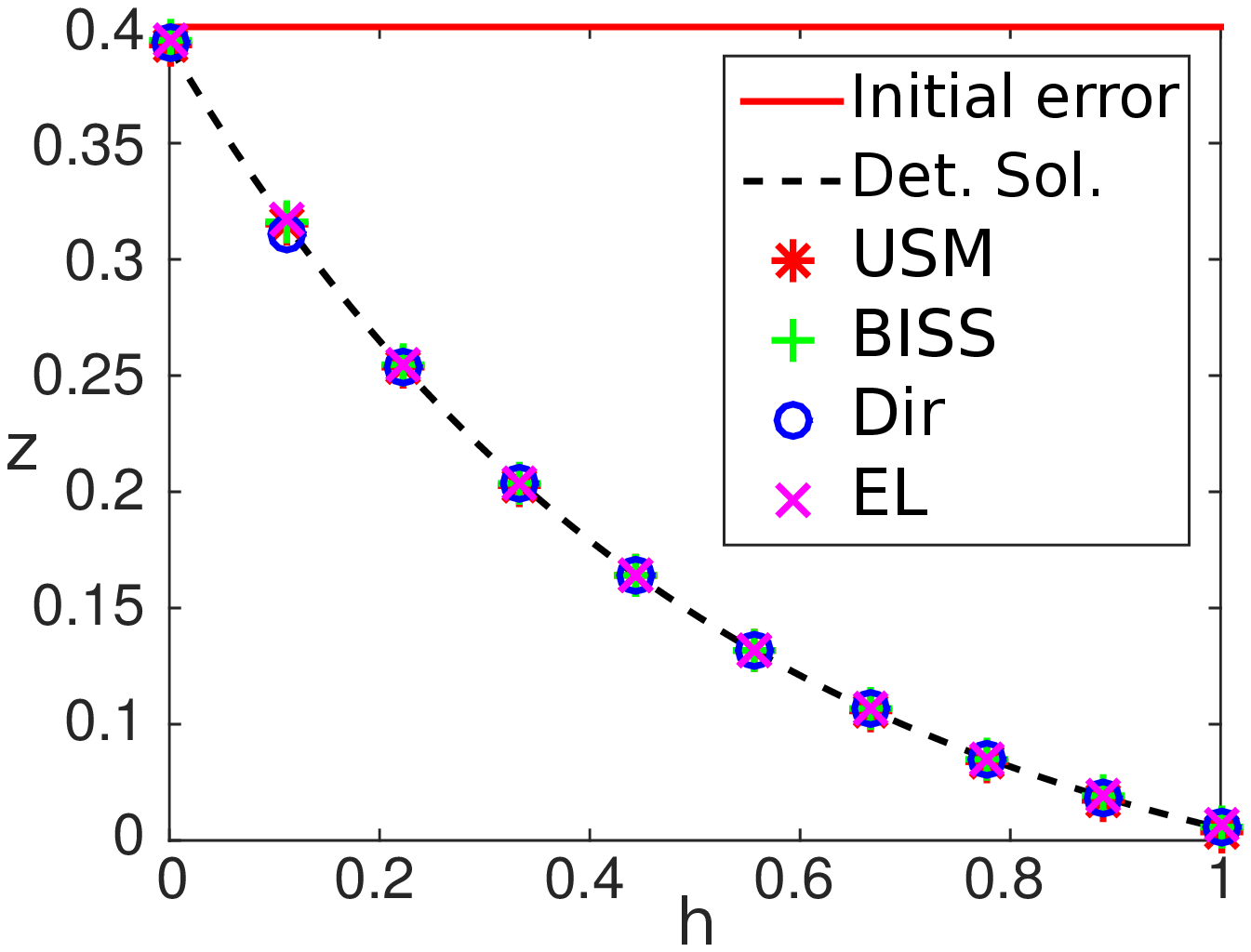}}
  \subfloat
  {\includegraphics[width=0.25\textwidth]{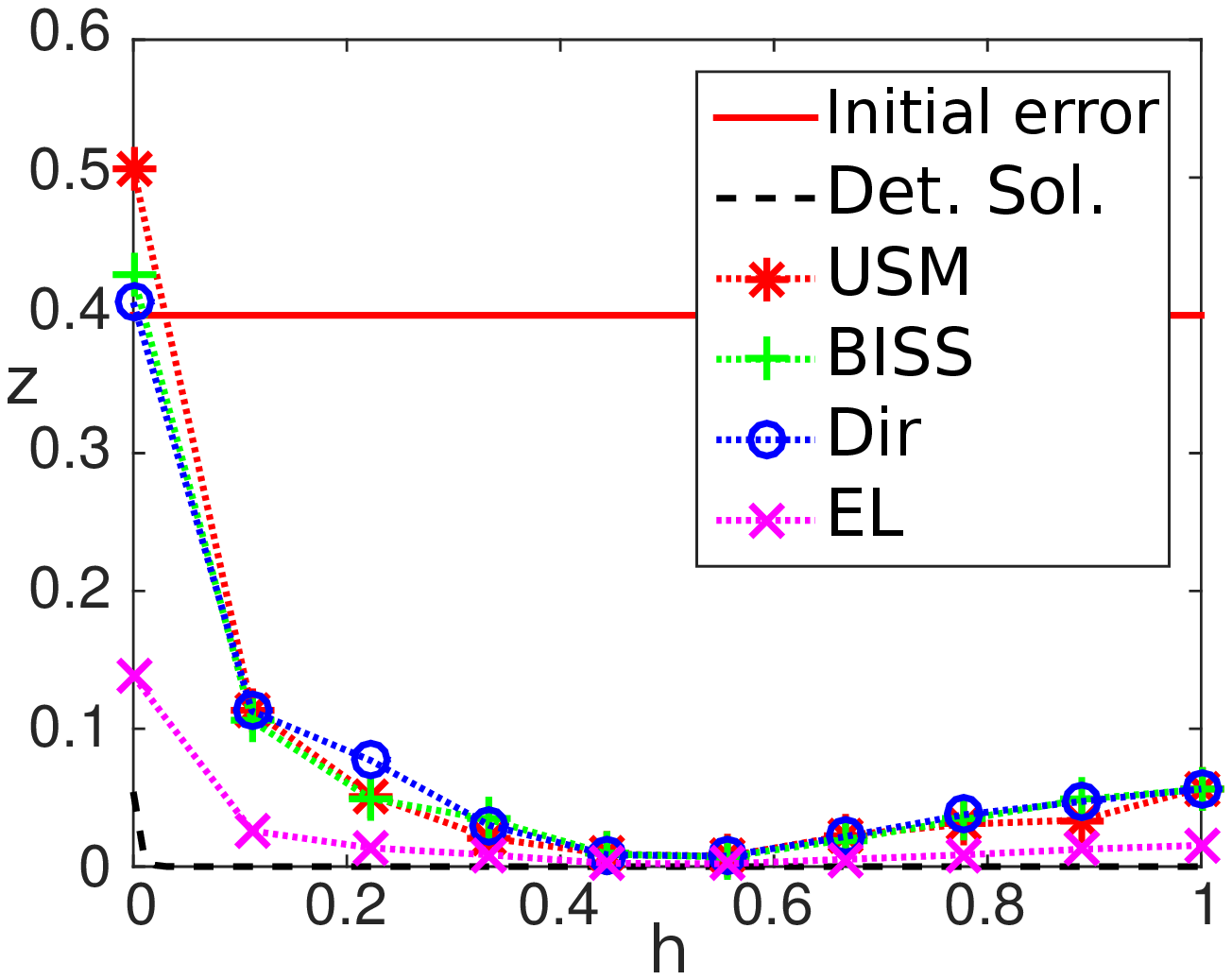}}
  \subfloat
   {\includegraphics[width=0.25\textwidth]{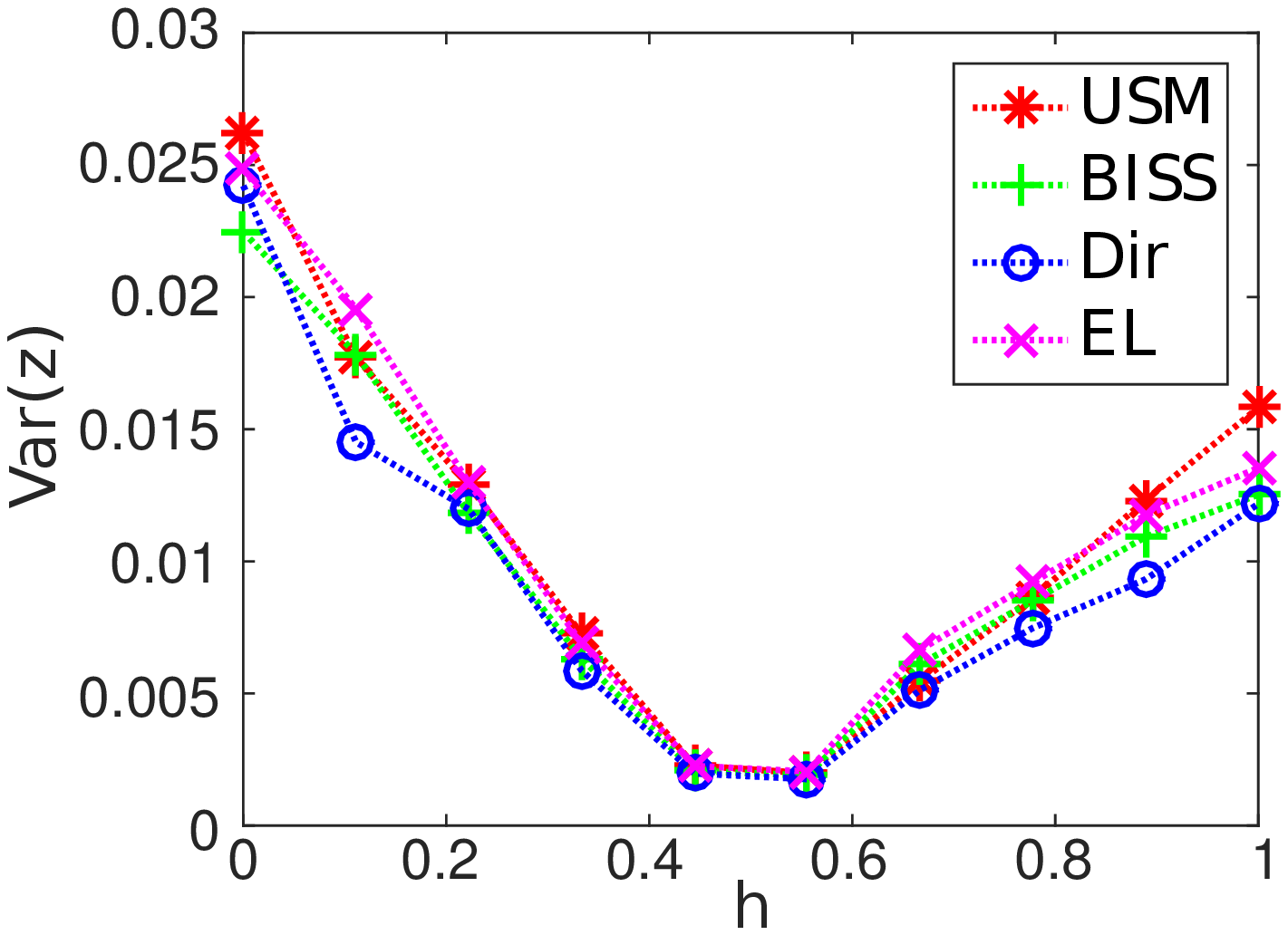}}
  \subfloat
  {\includegraphics[width=0.25\textwidth]{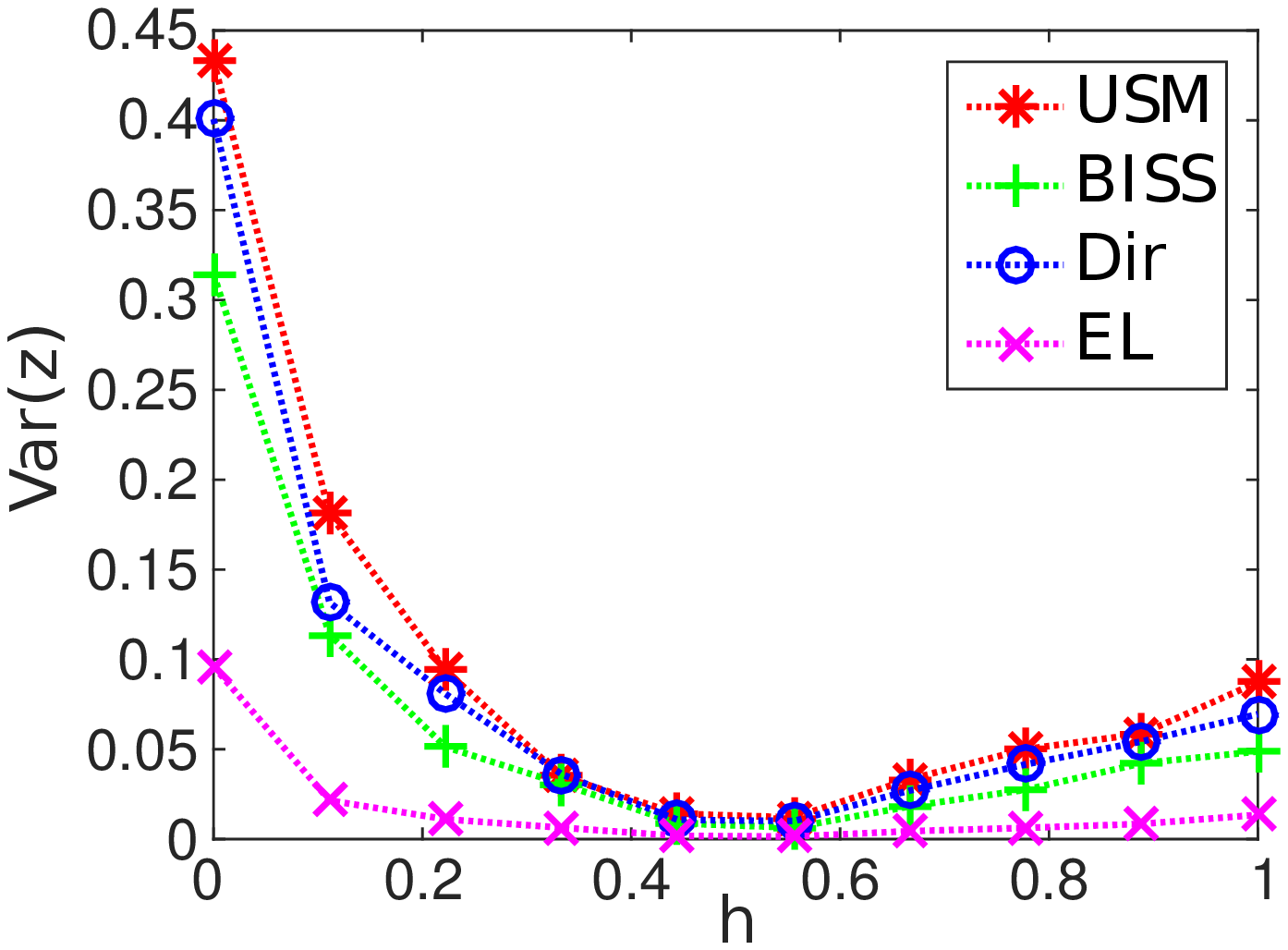}}
\caption{Comparison of models for the USM and continuous time limits. The red horizontal line represent the initial error $z(0)$. The dashed black line represents the solution of the deterministic limit. We display the average over 100 simulations and the corresponding variances. The continuous time limits have to be compared with the discrete USM model displayed in red stars. First panel: averaged $z$ at time $T=1$; second panel: averaged $z$ at time $T=100$; third panel: variance of $z$ at time $T=1$ and fourth panel: variance of $z$ at time $T=100$.
 }
\label{fig:NumExp}
\end{figure*}
We now perform some numerical experiments to validate the continuous time limits derived above. We consider a network of two connected agents $1$ and $2$ for simplicity. The probability $G^{(12)}$ that they interact is 1. This is the smallest network where interaction is possible. We also consider for simplicity the case of $2$ variants $V=2$. We compare the weak-noise SDE \eqref{eq:USMCTL} with the deterministic limit given by Eq.~\eqref{eq:USMCTL} in which the noise is neglected. To obtain a better insight in the dynamics, we consider the difference between the idiolects of the two agents, that is, we consider the variable $\vec z := \vec x^{(1)}-\vec x^{(2)}$. The deterministic evolution of $\vec z$ is given by
\begin{equation}
\dot{\vec z} = ((1-2h)M-I)\vec z:=A\vec z.
\end{equation}

We choose the formulation of Eq.~\eqref{eq:USMCTL}, since the other ordering of sampling and biasing has been shown to be ill-posed.
The matrix $A:=((1-2h)M-I)$ is negative definite unless $h=0$ and $M=I$, in which case $A=0$ and the difference is conserved. The consequence of this equation is that the behaviour of the two agents will converge as soon as there is either mutations $M\neq I$ or interactions $h\neq 0$ or both. If $h= 0$ and $M\neq I$, the convergence between the two agents is driven by the self-monitoring process. In fact, if there is mutation, there exists a vector $\vec x$ that minimizes the self-monitoring term and every agent will converge towards this particular idiolect, since the mutation matrix is the same for all agents. If $h \neq 0$ and $M=I$, then it is the interaction process that drives the convergence between the two agents. The greater the parameter $h$, the faster the convergence.

For this simple case, we compare the behaviour of the weak-noise SDE, the Dirichlet approximation and the deterministic limit. For the weak-noise SDE, we consider two different implementations. As we discussed in Sec.~\ref{sec:approx}, the normal approximation does not ensure the utterance vector $\vec u$ is bounded. This leads to numerical difficulties and various numerical strategies have been proposed. We review them in App.~\ref{App:NumSim}. 

For the parameters used in the simulation, we used short utterances $L=2$, a symmetric mutation matrix $M$ defined as
\begin{equation}\label{eq:M}
M := \begin{bmatrix}
1-q&q\\
q&1-q
\end{bmatrix},
\end{equation}
where $q=0.001$ is a mutation parameter.
 The initial condition is set to $x_1^{(1)}(0) = 0.2$ and $x_1^{(1)}(0) = 0.6$. This gives an initial difference $z(0) = 0.4$. In the simulation $h$ is varied from $0$ to $1$ and the statistics is performed on $100$ trajectories for each values of $h$. The results are given in Fig.~\ref{fig:NumExp}.

In Fig.~\ref{fig:NumExp}, we display the results for the different algorithm for a short time $T=1$ and for a long time $T=100$, where $T:= 2^8\delta t$ for continuous time limits and $\lambda=\delta t$ for the USM model. Changing the value of $\lambda$ in the USM, therefore, changes the timescale of the problem.
Results are displayed in Fig.~\ref{fig:NumExp}.
  For $T=1$, we observe that all the different algorithms agree well with the deterministic limit as shown in the first and the third panel of Fig.~\ref{fig:NumExp}. For longer times, however, the deterministic limit no longer agrees with the USM and its limits as shown in the second panel of Fig.~\ref{fig:NumExp}. This is due to the fact that after a long time, the deterministic part of \eqref{eq:USMCTL} tends to $0$ and the noise starts to contribute significanlty to the dynamics. This is a numerical justification that the noise term has to be kept. The target curve in the second panel of Fig.~\ref{fig:NumExp} corresponds to the USM discrete solution displayed as red stars. We see that the Dirichlet approximation and the \emph{backward implicit split step} (BISS)  implementation, see \cite{DangerfieldKayMacNamaraBurrage12} and App.~\ref{App:NumSim}, agree well with the discrete USM, but the \emph{explicit Euler} (EL) algorithm fails to capture the dynamics. The introduction of a control function that modifies the normal approximation leads to a better approximation. Therefore, we will use this algorithm for other numerical experiments. 

Note that the variance of all models vanishes for $h=0.5$, since in this case the dynamics of the variable $\vec z$ is always deterministic. For small values of $h$, the coupling is weak between the two agents and, as a consequence, the variance is larger for small values of $h$ than for high values of $h$. The variance of the BISS algorithm slightly underestimates the variance of the USM and Dirichlet approximation. This is a feature of this approximation and a consequence of the chosen control function given by Eq.~\eqref{eq:control}. 

These numerical simulations show that the noise term has to be kept to accurately capture the behaviour of the discrete USM. Recall that the deterministic limit corresponds to the KM expansion with the scaling $\lambda = \delta t$. The influence of the weak-noise has to be kept and the KM analysis is unsufficient to capture this dynamics.

In the next section, we discuss the coarse-graining procedure and explain how to obtain a stochastic heterogeneous mean field approximation of the USM.

\section{\label{sec:HMF}Heterogeneous mean field}
The main result of this paper is the derivation of a coarse-grained approximation of the USM in the form of a
stochastic heterogeneous mean field (sHMF) approximation, which is based on the idea that the behaviour of the complete network can be approximated by a smaller network of classes of agents grouped according to a relevant property. The sHMF we present in this paper is based on grouping by degree, similarly to what is done in \cite{newman2002spread}, but other grouping choice can be made. This grouping technique allows a coarse-graining procedure and the time scale of the approximation obtained is $t_{\mathcal G}$ instead of $t_{\rm int}$, that is, we obtain an approximation at the population level, thus realizing the diagonal arrow of Fig.~\ref{fig:CoarseContinuous}.

The main advantage of this approach is to keep the stochasticity of the model, while throwing away a lot of the network structure. The approximation obtained takes the form of a system of SDEs capturing the behaviour of the entire agent-based model. With this approximation, the influence of the different parameters on the population behaviour can be analyzed. 
In the rest of this section, we first discuss the network  and the state space reduction induced by a HMF approach, derive the sHMF of the USM and apply it to simple network topologies. We leave the discussion of complicated topologies for a further paper and focus in this paper on regular and star-shaped networks. 
In the case of regular networks, there is a single class of nodes and the sHMF reduces the dynamics to a single SDE. This SDE is of the same form as a WF diffusion process, see App.~\ref{App:USM}, and known results about this process can be applied. We also compare trajectories of the discrete USM with those of the sHMF to qualitatively validate the approximation. Unfortunately, it is not possible to provide a good analysis of pathwise convergence of the sHMF to the USM, since the sources of noise are of different natures. We then discuss the results for star-shaped network. This example illustrates the robustness of the sHMF for a very heterogeneous network.

\subsection{\label{ssec:GraphReduc}Graph and state space reduction}
\begin{figure*}[ht]
\subfloat
{
\includegraphics{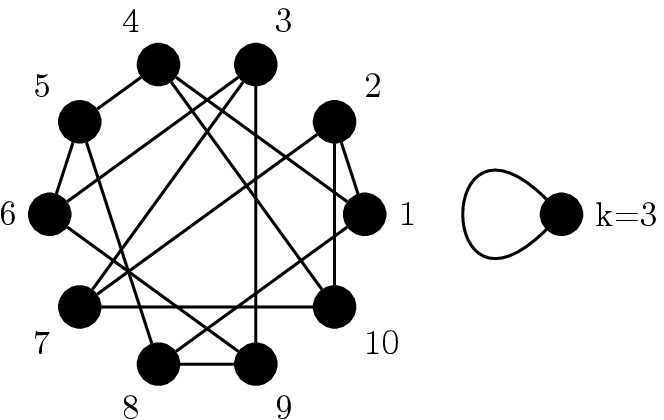}
}$\qquad$
\subfloat
{
\includegraphics{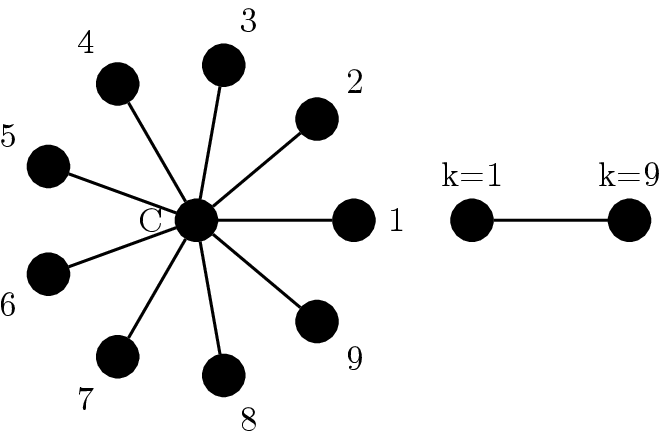}
}
\caption{Illustration of the network reduction for regular networks (left panel) and star-shaped networks (right panel). The left part is the original network and the right part is the reduced network.}
\label{fig:ReducedNetwork}
\end{figure*}
We now describe the graph and state space reduction induced by an sHMF approximation. The idea is to group the nodes according the a relevant property. This partition of the nodes in classes implies the existence of an equivalence relation, where the element of the node partition are equivalence classes. In this paper, we group the nodes by degree. This grouping is common in HMF approximations, see for example \cite{newman2002spread}. Note that other groupings are possible; one can group all the nodes and obtain a mean field approximation, or one can group nodes by communities. In each case, a partition in equivalence classes is implied.

In this paper, we group the nodes by degree, that is, we introduce the equivalence relation $\sim_{\rm deg}$ defined as
\[
i \sim_{\rm deg} j \text{ if } {\rm deg}(i) =  {\rm deg}(j),
\]
where ${\rm deg}(i)$ is the  degree of node $i$. Let ${\rm deg}(i)=k$, we then denote the corresponding equivalence class as $[k]$.
The nodes of the reduced graph are given by the classes $[k]$ and are given a weight $N_k$ representing the number of nodes contributing to the class $[k]$. Links between degree classes $[k]$ and $[k']$ exist whenever there is a link connected a node of degree $k$ to a node of degree $k'$ in the original network. These directed links are weighted by $p(k'|k)$, which represents the probability that a node of degree $k$ is connected to a node of degree $k'$. Note that in general $p(k'|k)\neq p(k|k')$ and that self-links are possible, since different nodes of same degree can be connected together.
\begin{example}For example, the reduced graph of a regular network (a network in which all nodes have the same degree) is a single node with a connection to itself, see left panel of Fig.~\ref{fig:ReducedNetwork}. The reduced graph of a star-shaped network has two connected nodes, but no connection to itself, since in this topology the node of one class always interact with nodes of the other class, see right panel of Fig.~\ref{fig:ReducedNetwork}.
\end{example}

In the sHMF, each  degree classes is described by a single belief distribution $\vec x^{(k)} \in \mathds P_V$ defined as 
\[
\vec x^{(k)} := \frac1{N_k}\sum_{i\in [k]}\vec x^{(i)},\quad \vec x^{(k)} \in \mathds P_V,
\] 
where $N_k$ is the number of agents of degree $k$ in the network.
 If there are $K$ classes, the dimension of the state space is $K(V-1)$, since $\mathds P_V$ is of dimension $V-1$ because of the normalization constraint. In the original model, the dimension of the state space is $N(V-1)$. If $K\ll N$, the sHMF significantly reduces the dimension of the state space.

\subsection{\label{ssec:DerivHMF}Derivation of the stochastic Heterogeneous Mean Field approximation}

We can now derive the sHMF of the USM. This is where the work done in previous sections, and in particular the continuous in $L$ normal approximation of Sec.~\ref{sec:approx}, pays off. 
The sHMF uses a time unit corresponding to the network time $t_{\mathcal G}$. The core idea of the approximation is to consider a class of nodes as a single agent, which corresponds to the vertical arrow between Disc. ABM and Disc. PBM in  Fig.~\ref{fig:CoarseContinuous}, and to use the new continuous time limit obtained in the previous section to implement the horizontal arrow between Disc. PBM and Cont. PBM in  Fig.~\ref{fig:CoarseContinuous}. To do so, we  
group all the agents belonging to the same degree class and ask them to produce all the utterances they have to utter during a complete network update and consider the results as a single class utterance of length $L_k := \frac{kL}{E}N_k$. At each network time step, all degree classes exchange their class utterance with the other degree classes. A weight $p(k'|k)$ is given to these utterances, proportional to the probability that the two degree classes are connected. Since $L_k$ is usually large, the normal approximation, which fails in the two agents case, is now justified by the central limit theorem and one can use it to approximate the average utterance by:
\begin{equation}\label{eq:HMFutterance}
\vec u^{(k)} = M\left({\vec x}^{(k)} + \frac{1}{\sqrt{L_k}}D(\vec x^{(k)})\boldsymbol\xi^{(k)}\right),
\end{equation}
where $M$ is the production error, or mutation, matrix and $D(\vec x)$ is a Cholesky square root of the covariance matrix of a multinomial distribution, see App.~\ref{App:WF}, and $\boldsymbol\xi^{(k)}\sim\mathcal N(0,\vec I)$ is a normally distributed random vector.

Assuming that $G^{(ij)} = \frac1{E}\delta_{i \leftrightarrow j}$, that is, the probability to pick an edge is uniform. The averaged change of a degree class $[k]$ at the network level is given by
\begin{equation}\begin{aligned}\label{eq:HMFdx}
\delta \vec x^{(k)} =&\ \lambda\frac{(1-h)k}{E}(\vec u^{(k)}-\vec x^{(k)}) \\&+ \lambda \frac{hk}{E}\sum_{k'}p(k'|k)(\vec u^{(k')}-\vec x^{(k)}),
\end{aligned}\end{equation}
where $p(k'|k)$ is the probability that a node of degree $k$ is connected to a node of degree $k'$ and $E$ is the number of edges of the network.

Introducing the degree $k$ utterance \eqref{eq:HMFutterance} into Eq.~\eqref{eq:HMFdx} and introducing the scaling $dt = \frac1{E}$, which is motivated by the fact that the interaction time is much faster than the network time, see Fig.~\ref{fig:timescales}, gives the following SDE
\begin{equation}\begin{aligned}\label{eq:USM-sHMF}
&d\vec x^{(k)} = \lambda\Bigg[(1-h)k(M{\vec x}^{(k)}-\vec x^{(k)}) \\&+ hk\sum_{k'}p(k'|k)(M{\vec x}^{(k')}-\vec x^{(k)})\Bigg]dt\\
&+\lambda\Bigg[(1-h)\sqrt{\frac{k}{LN_k}}MD(\vec x^{(k)})d\boldsymbol W_t^{(k)}\!\!\\&+hk\!\sum_{k'}p(k'|k)\frac{1}{\sqrt{Lk'N_{k'}}}MD(\vec x^{(k')})d\boldsymbol W_t^{(k')}\Bigg],
\end{aligned}\end{equation}
where the time is measured in $t_{\mathcal G}$ units.
Eq.~\eqref{eq:USM-sHMF} is the continuous time sHMF approximation of the USM. The first two terms describe the influence of the self-monitoring and accommodation processes and the last two terms model the corresponding noises. There is one such equation for each degree class $[k]$.

This approximation greatly reduces the number of degrees of freedom whenever $K\ll N$, where $K$ is the number of equivalence classes $[k]$. The number of agents $N_k$ in a class $[k]$ only enters Eq.~\eqref{eq:USM-sHMF} as a parameter of the noise coefficients. The noises are therefore dependent of the size of the network. For large networks, the contribution of the noise is small and vanishes in the limit of infinite networks. In other words, the global stochastic dynamics of the model is a finite size effect. The parameter $L$ also controls the amplitude of the noise. The shorter the utterance, the larger the noise. This justifies the interpretation of $L$ as describing the variability of a speaker, see Tab.~\ref{tab:USMparam}. 

In the sHMF, we are throwing away a lot of information about the topology of the network, conserving only the different degree classes. If we model the social interaction by randomly ordering the edges and going through them exactly once at each network time, the nodes with a large number of neighbours interact more often than nodes with a small number of neighbours. As a result, we expect the evolution of the different classes of nodes to evolve on a different time scale. In Eq.~\eqref{eq:USM-sHMF}, the time scale difference is encoded in the dependency on $k$ of the dynamics of $\vec x^{(k)}$.

We expect the sHMF to be a good approximation if the number of agents in each degree classes is sufficiently large for the normal approximation to hold and if the nodes forming a class are well-connected. Both of these conditions are satisfied for regular networks. A limiting case is given by star-shaped networks, in which there is no direct connections between nodes of degree $1$ and where there is a single node of degree $N-1$. In this case, both conditions are violated and we show that the sHMF nevertheless captures well the dynamics of the system.

In the following, we apply the sHMF to regular networks and to star-shaped networks. The regular network analysis allows us to study in detail the influence of the different parameters and the star-shaped network illustrates the robustness of the method.

\subsection{\label{ssec:RegNet}Regular Networks and Wright-Fisher SDE}
The case of regular networks is particularly interesting, since its sHMF takes the form of a Wright-Fisher diffusion, which has been widely studied, much is known about the behaviour of this process and we can apply this knowledge to the study of the sHMF of the USM. The left panel of Fig.~\ref{fig:ReducedNetwork} illustrates the type of network we are considering, together with the reduced network of degree class on which the sHMF is defined.

For simplicity, we restrict the discussion to the case of two variants $V=2$ and we choose of mutation matrix $M$  of the form \eqref{eq:M}, with $q=10^{-3}.$
The Cholesky square root $D(\vec x)$ is given by Eq.~\eqref{eq:2V-D}. Under these assumptions, the sHMF of the regular network is given by
\begin{equation}\begin{aligned}\label{eq:HMFRegNet}
dx_1^{(k)} &= k\lambda({x'}_1^{(k)}-x^{(k)}_1) dt\\&\quad +\lambda(1-2q)\sqrt{\frac{k}{LN}}\sqrt{{x}_1^{(k)}(1-{x}_1^{(k)})}dW_t^{(k)}\\
& = -\gamma(x^{(k)}_1-\frac{1}{2}) dt+\sigma\sqrt{{x}_1^{(k)}(1-{x}_1^{(k)})}dW_t^{(k)},
\end{aligned}\end{equation}
where $\vec x' = M\vec x$ and $x^{(k)}_2= 1-x^{(k)}_1$ to conserve probability. We also introduced $\gamma = 2qk\lambda$ and $\sigma =\lambda (1-2q)\sqrt{\frac{k}{LN}}$. The time has to be measured in $t_{\mathcal G}$ units.

In order to simplify the discussion, we scale the time variable as $t':= \lambda k t_{\mathcal G}$. With this scaling, Eq.~\eqref{eq:HMFRegNet} can be rewritten as
\begin{equation}\label{eq:HMFRegNet.2}
dx_1^{(k)} =  -\gamma'(x^{(k)}_1-\frac{1}{2}) dt'+\sigma'\sqrt{{x}_1^{(k)}(1-{x}_1^{(k)})}dW_{t'}^{(k)},
\end{equation}
where
\begin{equation}
\begin{aligned}
\gamma'&=2q,\\
\sigma'&=(1-2q)\sqrt{\frac{\lambda}{LN}}.
\end{aligned}
\end{equation}

Eq.~\eqref{eq:HMFRegNet.2} is a WF process, as discussed in App.~\ref{App:WF}. This process occurs in many different contexts such as population genetics and economics, see for example \cite{BouleauChorro15}. The type of noise occuring in Eq.~\eqref{eq:HMFRegNet.2} can be found in another model for language change in which an age-structured population is considered, see \cite{Mitchener09}.

The three relevant parameters controlling the dynamics are $\lambda k$, $q$ and $r:=\frac{\lambda}{LN}$.
The time scale evolution is controlled by the product $\lambda k$ of the learning rate and of the degree of the class. This is expected, since $\lambda$ models the amplitude of change at each time step and since an agent of degree $k$ interacts $k$ times during a single network update.
The parameter $q$ models the influence of error production and innovations. If $q=0$, then there is no error and no innovation. In this case, $\lambda'=0$ and $\sigma' = \sqrt{r}$. In other words, the dynamics is only driven by noise and the boundaries are absorbing. Once the population reaches a consensus, the state of the system no longer changes. The other extreme case is when $q= \frac12$. In this case, the multiplication by $M$ in \eqref{eq:HMFutterance} randomizes the output and the noise information is lost. In this case, the noise coefficient $\sigma'$ vanishes and the dynamics is driven by the drift term and the solution deterministically goes to $x_1^{(k)}= \frac12$.
The parameter $r$ controls the size of the noise and is proportional to $\lambda$, and inversely proportional to $N$ and $L$. When $r$ is large the noise dominates the dynamics and the solution is pushed towards the boundary of the domain and when $r$ is small the drift term dominates the dynamics and the solution is pushed towards the centre of the domain. Therefore, we expect a change of the stationary distribution shape between a U-shaped and a bell-shaped distribution by varying $r$ and $q$.

\begin{figure}[t]
\centering
\includegraphics[width=0.4\textwidth]{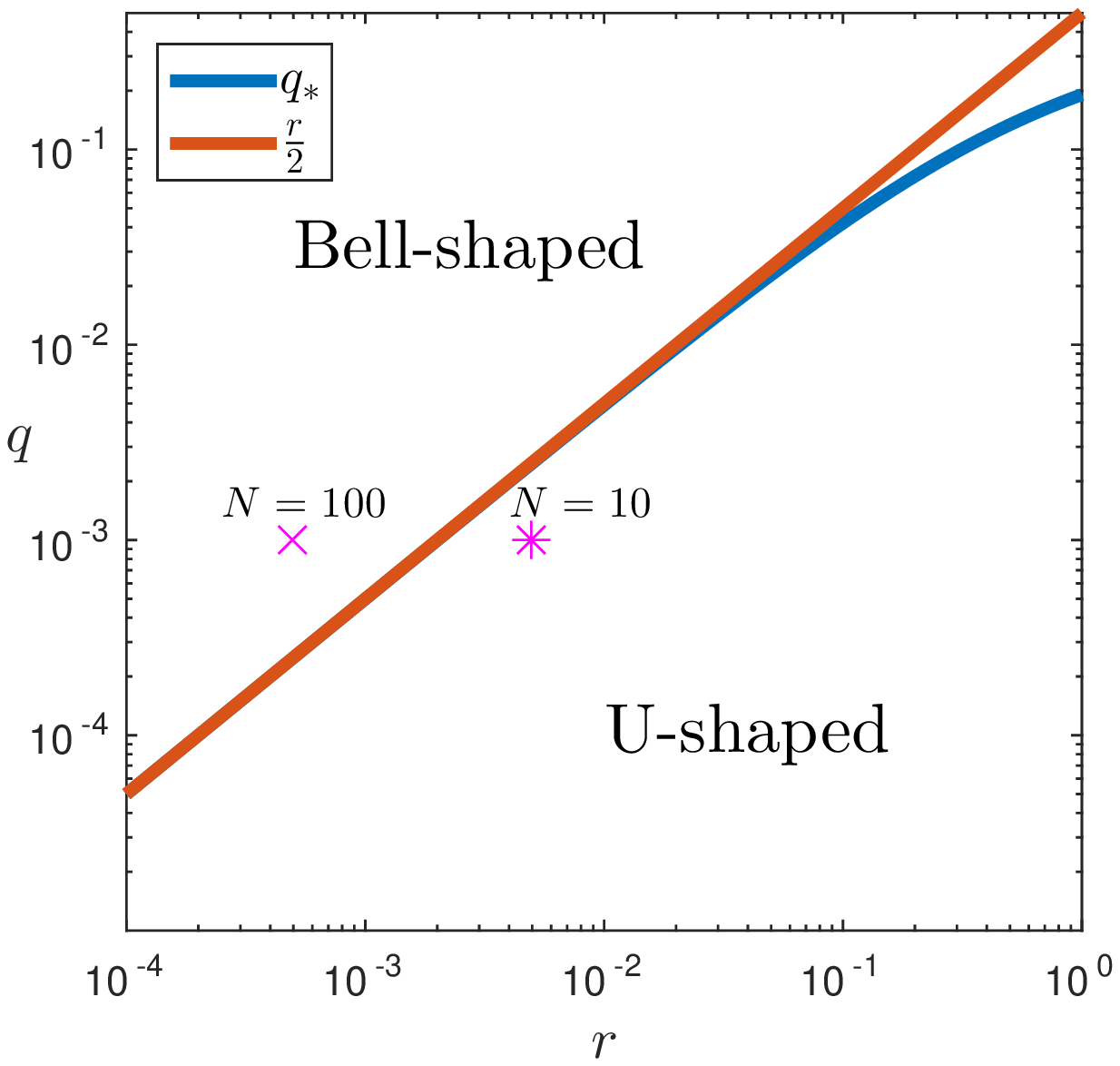}
\caption{Illustration of the critical parameter $q_*(r)$ separating the Bell-shaped and the U-shaped domain. The red curve is the approximate behaviour. For illustration, we display the positions of the two examples considered in this section (regular network of degree $3$ for $N=10$ and $N=100$ agents).}
\label{fig:q}
\end{figure}

 We can now take advantage of the WF form of the sHMF of the regular network for which the stationary distribution is known and takes the form of a Beta distribution, see for example \cite{BaxterEtAl06}. For long times, the probability $p_*(x)$ that a trajectory reaches a certain value $x$ is given by
 \begin{equation}\label{eq:StatDistrib}
p_*(x) = 2\frac{\Gamma\left(\frac{\gamma}{\sigma^2}+\frac1{2}\right)}{\Gamma\left(\frac{\gamma}{\sigma^2}\right)+\Gamma\left(\frac1{2}\right)}(4x(1-x))^{(\frac{\gamma}{\sigma^2}-1)}.
\end{equation}
For more than two variants, one can generalize this formula. The resulting Dirichlet distribution can be found in \cite{Baxter07}.
In our case, the single parameter of this distribution is given by
\begin{equation}\label{eq:criticalCond}
\frac{\gamma}{\sigma^2} = \frac{2q}{(1-2q)^2r}.
\end{equation}
We see that this parameter only depends on $q$ and $r$, as expected. The distribution \eqref{eq:StatDistrib} changes from a bell-shaped distribution for $\sigma^2>\gamma$ to a U-shaped distribution for $\sigma^2<\gamma$, with a transition when $\sigma^2=\gamma$. In the bell-shaped regime, there is no convention emerging and the agents are probability matching and the dynamics is dominated by the deterministic term. In the U-shaped regime, conventions emerge, but are not stable unless $q=0$, in which case the distribution degenerates to the discrete probability mass function weighting only $x=0$ and $x=1$. From Eq.~\eqref{eq:criticalCond}, one obtains the critical value for $q_*(r)$ given by
\begin{equation}
q_*(r) = \frac{r}{1+2r+\sqrt{1+4r}},
\end{equation}
which behaves as $q_*(r) \propto \frac{r}{2}$, when $r\to 0$.

Fig.~\ref{fig:q} summarizes the behaviour of regular graphs. The exact critical value $q_*$ and its asymptotic behaviour are displayed, separating the parameter space into regions of  U-shaped and bell-shaped stationary distributions. Since the parameter $r$ is inversely proportional to $N$, this phase diagram is the signature of a finite size effect. For an infinite graph, the distribution is always bell-shaped and no convention can ever globally emerge.

For a fixed parameter $q$, decreasing the parameter $r$ leads to a phase transition from a U-shaped to a bell-shaped distribution. We recall that $r$ is proportional to $\lambda$ and inversely proportional to $N$ and $L$.

\begin{figure}[t]
  \centering
  \subfloat
  {\includegraphics[scale=0.3]{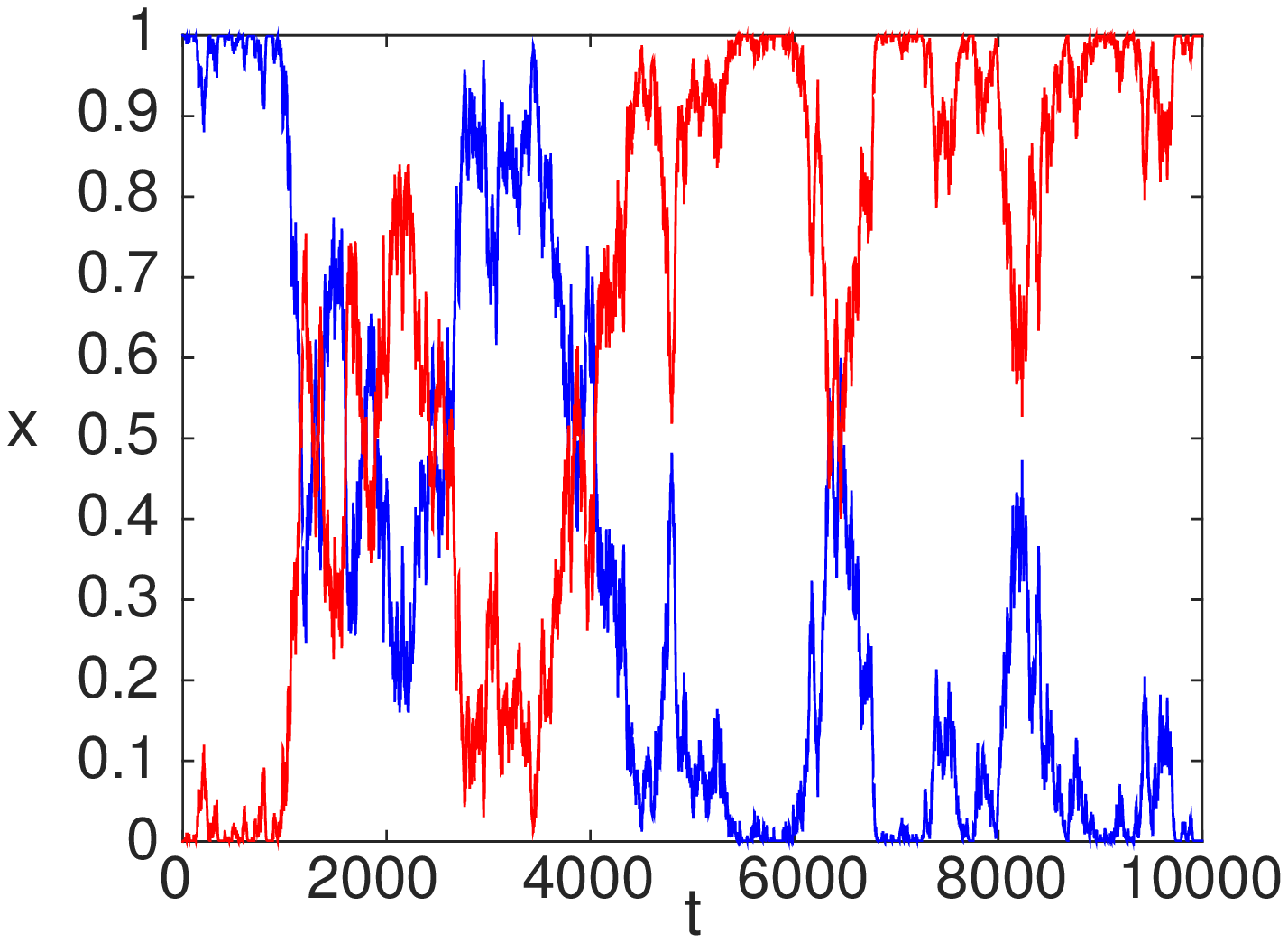}}
  \subfloat
  {\includegraphics[scale=0.3]{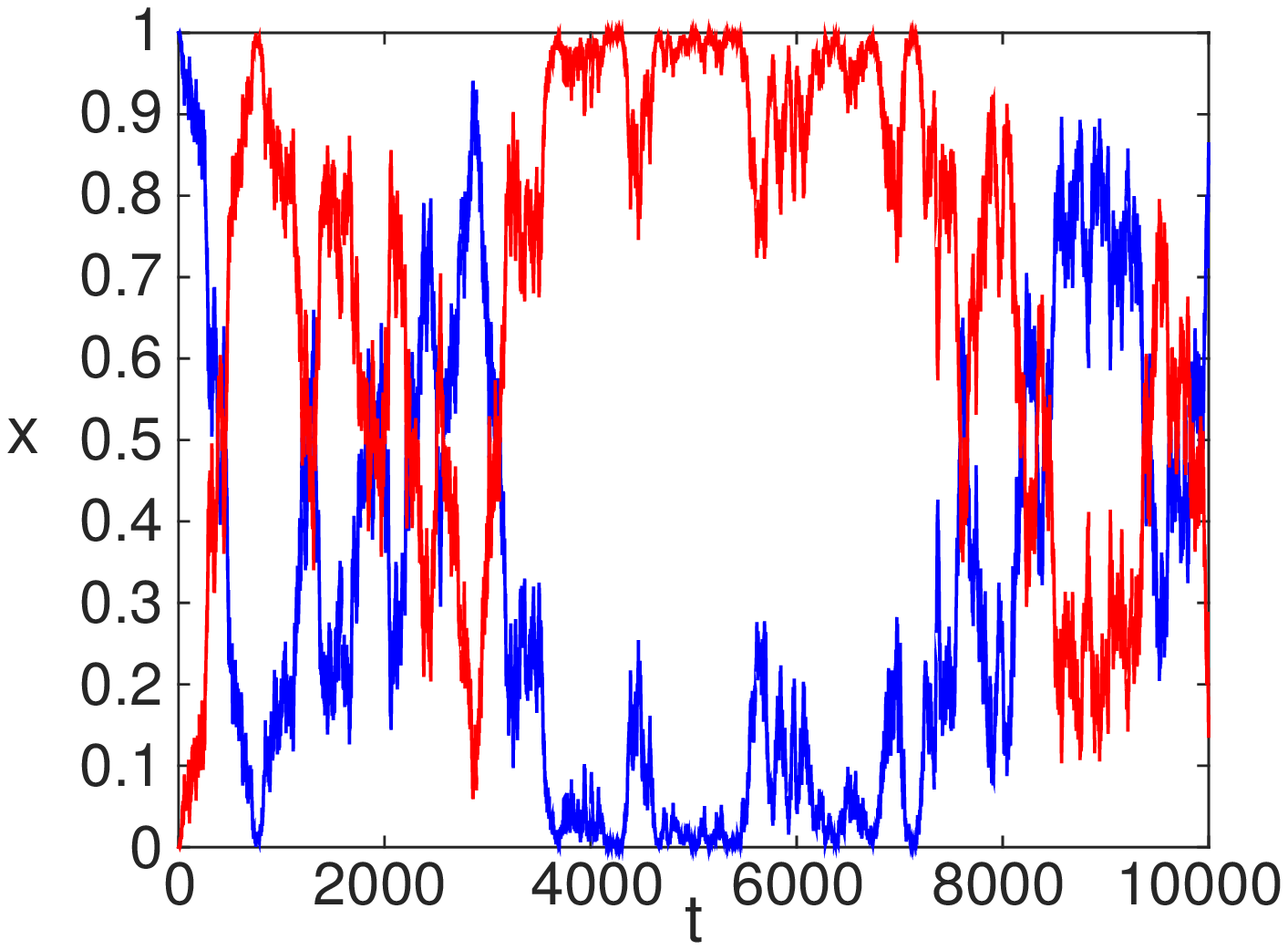}}
  \hspace{5pt}
  \subfloat
  {\includegraphics[scale=0.3]{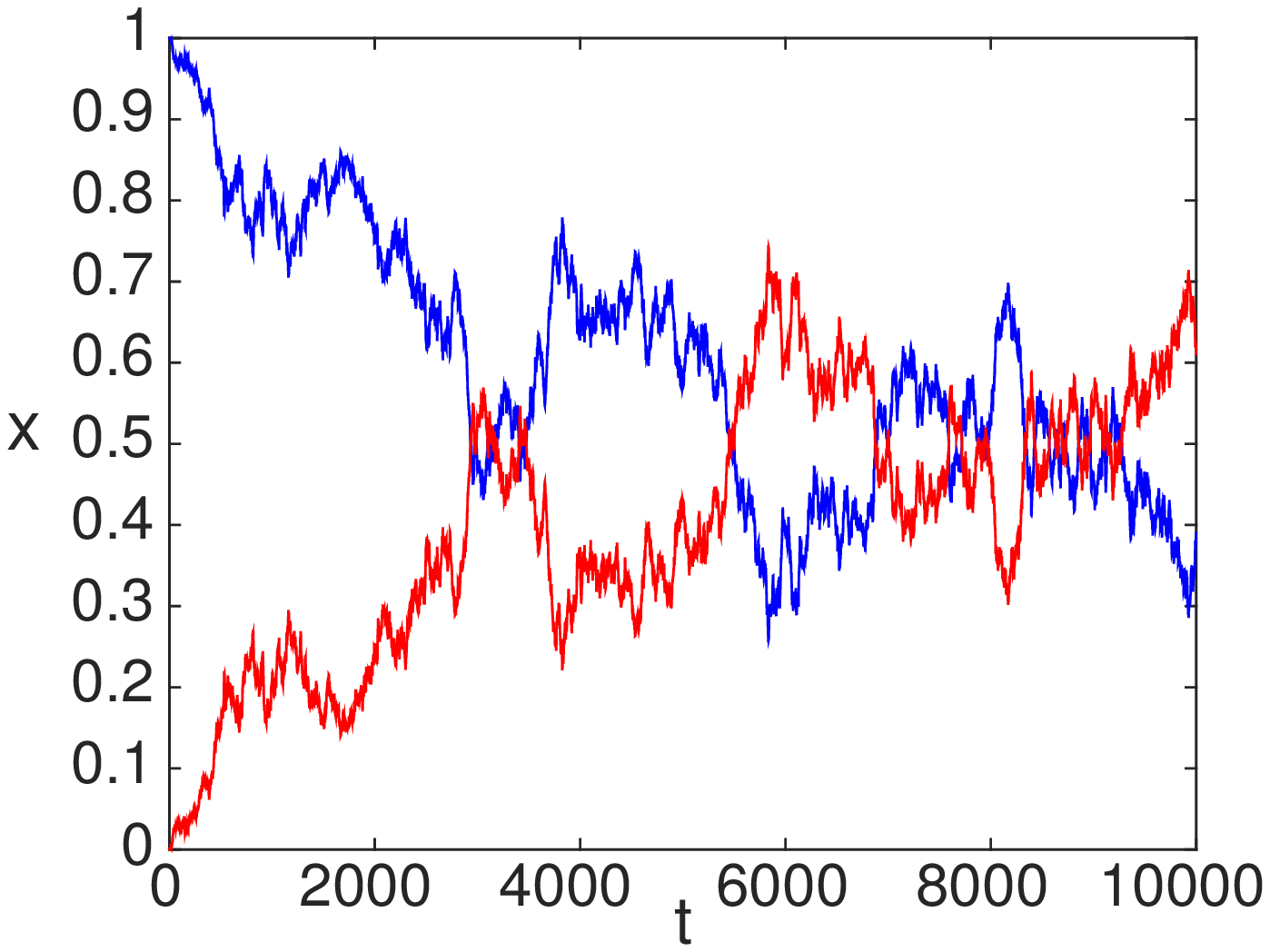}}
  \subfloat
  {\includegraphics[scale=0.3]{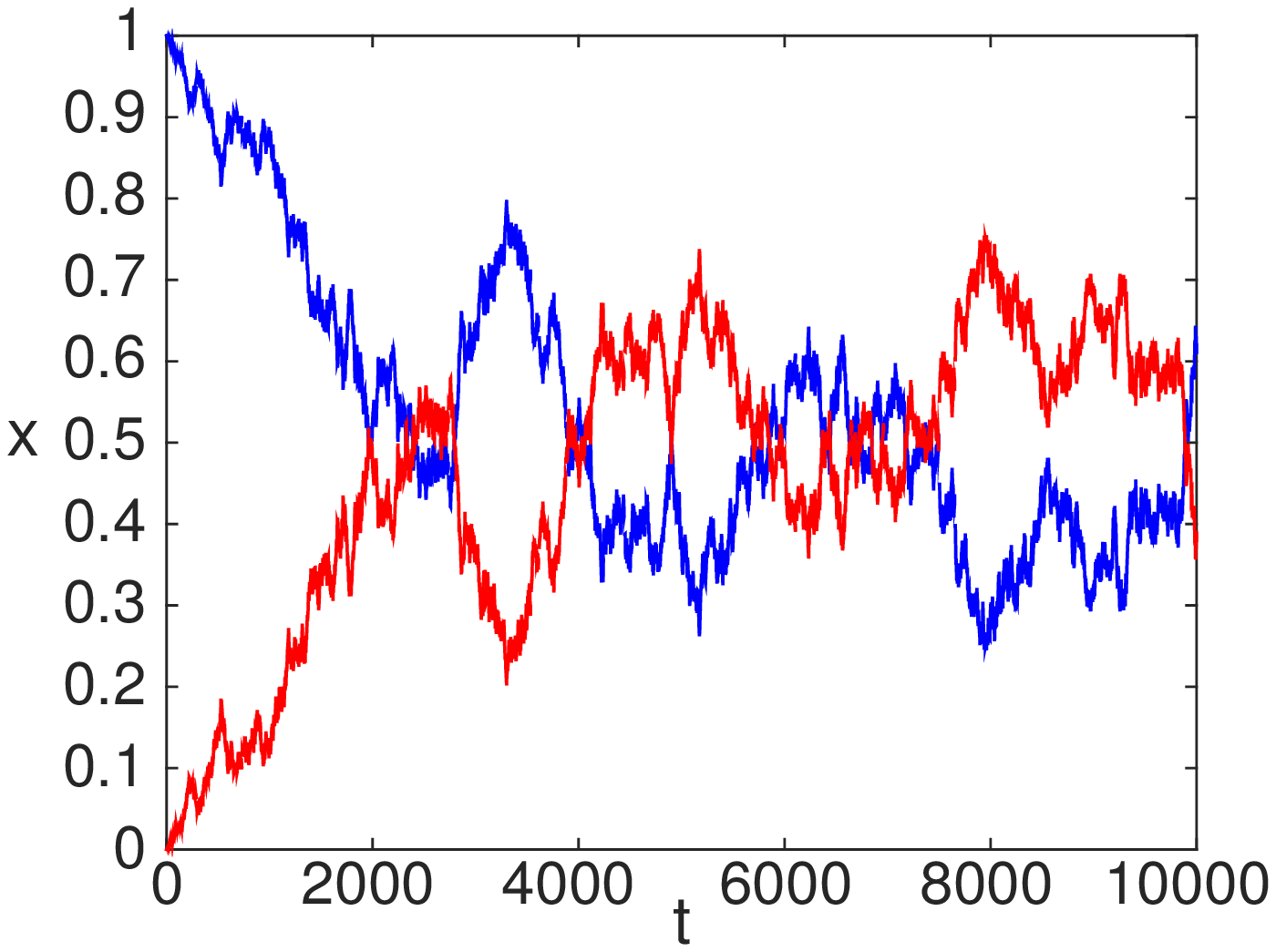}}
     \caption{Comparison between the discrete USM model (first column) and the sHMF limit of it (second column). For these simulations, the paramaters are $h = 0.5$, $\lambda =0.1$, $V=2$, $L=2$, $T=10^4$ and $q= 0.001$ and the number of agents in $10$ in the first row and $100$ in the second row. The regular graph is of degree $k=3$. At the beginning of the simulation, all agents share the convention to use variant $v=1$.}
  \label{fig:RegNetUSMsHMF}
\end{figure}

The parameter $k$, representing the degree of the regular graph, only contributes to the $\lambda k$ time scale parameter and, therefore, has no influence on the shape of the stationary distribution of the averaged system. 

For the numerical simulations, we choose regular networks of degree $k=3$. The agents choose between $V=2$ variants and produce utterances of length $L=2$ for $T=10^4$ network updates. For the other parameters, we choose $h=0.5$, $q=0.001$. We then change the number of agents from $N=10$ to $N=100$, which corresponds to values of $r = 200$ and $r= 2000$, respectively. With these parameters, the critical values of the mutation parameter are $q_*\approx2.475\cdot10^{-3}$ and $q_*\approx2.498\cdot10^{-4}$. These values are plotted in Fig.~\ref{fig:q}. Since the chosen value of $q<q_*$ for $N=10$, we expect a U-shaped distribution and since $q>q_*$ for $N=100$, we expect a bell-shaped distribution.

Results for the trajectories of the discrete USM and for the corresponding sHMF are displayed in Fig.~\ref{fig:RegNetUSMsHMF}. The results of the sHMF are in good qualitative agreement with the results of the discrete USM model and can therefore be used to characterize the behaviour of the system.
\begin{figure}[t]
\centering
   \subfloat
   {\includegraphics[scale=0.3]{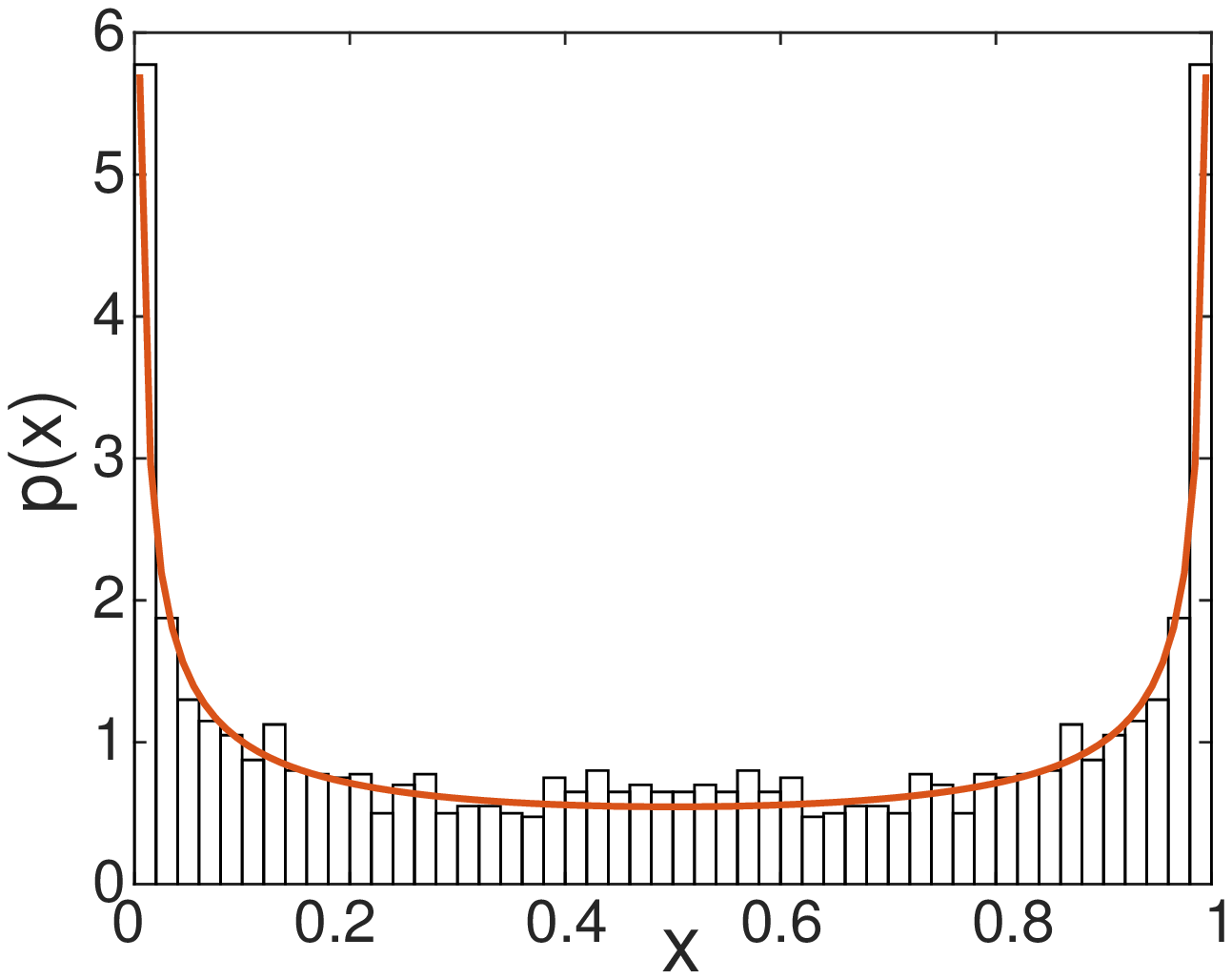}}
  \subfloat
  {\includegraphics[scale=0.3]{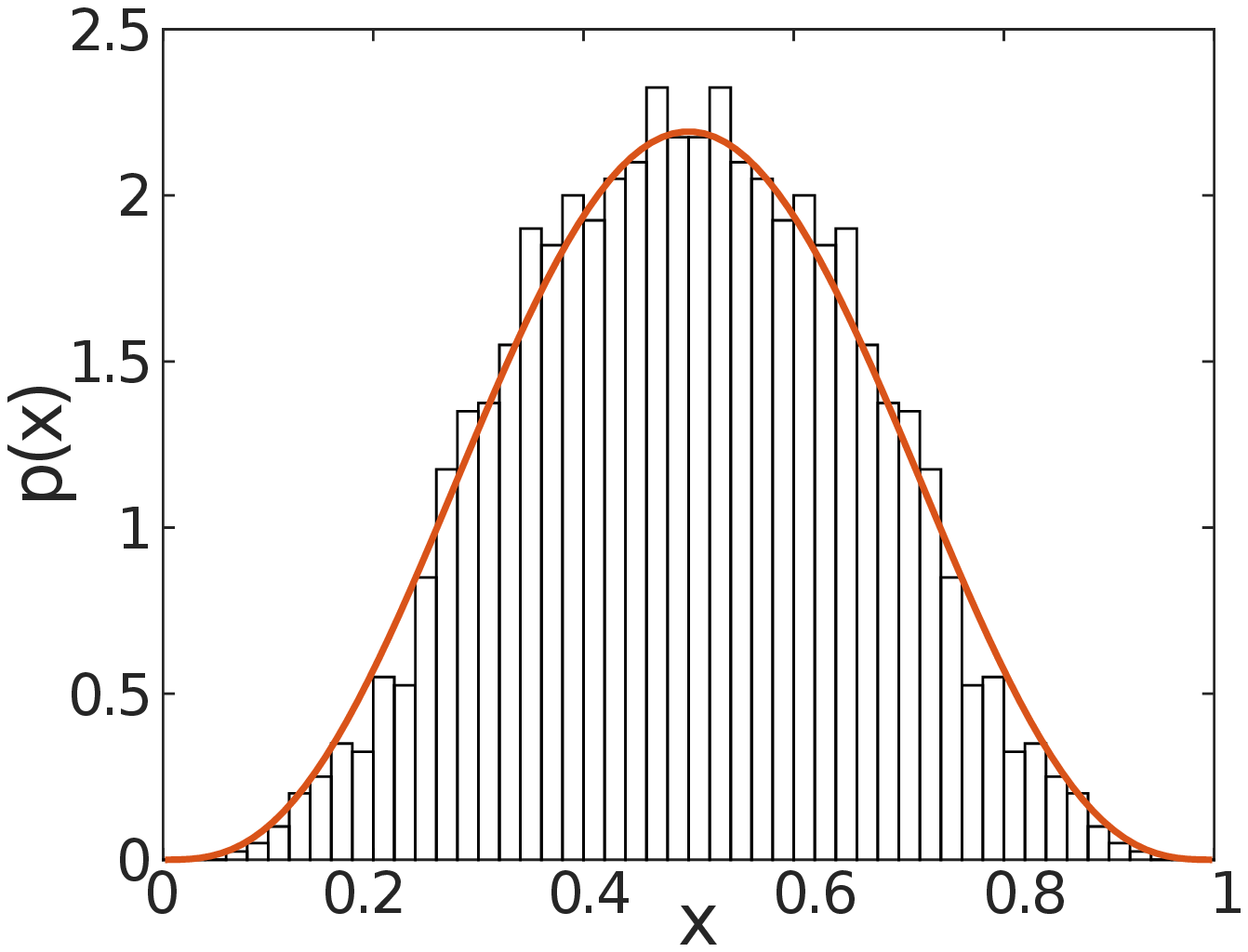}}
  \caption{Comparison between the stationary distribution of the discrete USM on a regular network with the distribution predicted by the sHMF. Left panel: stationary distribution for 10 agents. Right panel: stationary distribution for 100 agents.}
  \label{fig:RegNetDistrib}
\end{figure}

In order to validate the sHMF approximation of the USM, we computed the stationary distribution of the discrete USM and compared the results with the analytical prediction of its sHMF approximation. The results displayed in Fig.~\ref{fig:RegNetDistrib} are excellent already with a relatively small statistics of $1000$ trajectories. Since the computation of the stationary distribution of the discrete USM is time consuming and due to the symmetry of the dynamics, we augmented the statistics by considering both $x_1$ and $x_2$ at the end of the simulation, since the two variants are equivalent.

For regular networks, the parameter $h$ does not play a role in predicting the population-averaged stationary distribution. This has been verified by performing the simulation for different values of $h$ (not shown). However, in \cite{BaxterEtAl06} it is shown that $h$ does play a role on the marginal stationary distribution, in other words, $h$ has an influence on the stationary distribution of higher order moments, rather than on the stationary distribution of the averaged behaviour analysed here. In order to better understand the effect of $h$ on the population averaged stationary distribution, we now consider the case of star-shaped networks.

\subsection{\label{ssec:StarNet}Star-shaped Networks}
We now consider the case of a heterogeneous network, namely, the star-shaped network. This kind of network is characterized by two classes of nodes, a central node of degree $N-1$ and $N-1$ nodes of degree $1$ connected to it. The right panel of Fig.~\ref{fig:ReducedNetwork} illustrates this kind of network, together with the reduced network used in the sHMF approximation. 

For this kind of network, the sHMF is expected to fail to capture efficiently the dynamics. This is due to the fact that the normal approximation is not well justified for the central node labelled $C$ in the right panel of Fig.~\ref{fig:ReducedNetwork}. Furthermore, all the degree one nodes interact through the mediation of this poorely approximated node.

In order to simplify the notation, we introduce the quantities
\[\begin{aligned}
\sigma_1 &=\lambda(1-2q)\frac1{\sqrt{L(N-1)}} ,\quad \gamma_1 =2q\lambda ,\\ \quad \sigma_N &=\lambda(1-2q)\sqrt{\frac{(N-1)}{{L}}} ,\quad \gamma_N =2q\lambda(N-1),
\end{aligned}\]
and we have the relations $\gamma_N=(N-1)\gamma_1$ and $\sigma_N=(N-1)\sigma_1$. 
With this notation, the sHMF formulation of the USM for a star-shaped network of $N$ agents reads
\begin{widetext}
\begin{equation}\label{eq:sHMF-star}
\begin{aligned}
dx_1^{(1)} =& \left[\gamma_1(1-h)\left(\frac{1}{2}-x^{(1)}_1\right) +\lambda h\left({x'}_1^{(N-1)}-x_1^{(1)}\right)\right]dt\\
&\quad+(1-h)\sigma_1\sqrt{x^{(1)}_1(1-x^{(1)}_1)}dW^{(1)}_t +h\sigma_1\sqrt{x^{(N-1)}_1(1-x^{(N-1)}_1)}dW^{(N-1)}_t,\\
dx_1^{(N-1)} =&\left[ (1-h)\gamma_N\left(\frac{1}{2}-x^{(N-1)}_1\right) +\lambda h\left({x'}_1^{(1)}-x_1^{(N-1)}\right)\right]dt\\
&\quad+(1-h)\sigma_N\sqrt{x^{(N-1)}_1(1-x^{(N-1)}_1)}dW^{(N-1)}_t +h\sigma_N\sqrt{x^{(1)}_1(1-x^{(1)}_1)}dW^{(1)}_t,
\end{aligned}\end{equation}
\end{widetext}
where ${x'}_1^{(i)}$ is the first component of $M\vec x^{(i)}$, $i = 1,N-1$.

For Eq.~\eqref{eq:sHMF-star}, we do not have an analytical form for the stationary distribution of $x_1^{(1)}$ and $x_1^{(N-1)}$. However, the results obtained for the regular network case can be used to gain some insights for this problem. For example, we observe that the noise magnitude is much larger for the central node than for the other nodes. This is a consequence of the time scale difference between the two classes of nodes.

In order to illustrate the behaviour of the star-shaped network and, in particular, the influence of the $h$ parameter, we performed simulations of the star-shaped network for parameters similar to those used for the regular network case. We consider $V=2$ variants that are used to produce utterances of length $L=2$, the mutation parameter entering the symmetric mutation matrix $M$ is fixed to $q=10^{-3}$. The learning rate is $\lambda = 0.1$ and the simulation ends after $T=10^4$ network timesteps. For these parameters, we vary the number of agents: $N=10$ or $N=100$ and the parameter $h$: $h=0.9$ and $h=0.1$. In these settings, we compare the behaviour of the discrete USM with the behaviour of the corresponding sHMF approximation.

\begin{figure*}[t]
  \centering
  \subfloat
  {\includegraphics[scale=0.3]{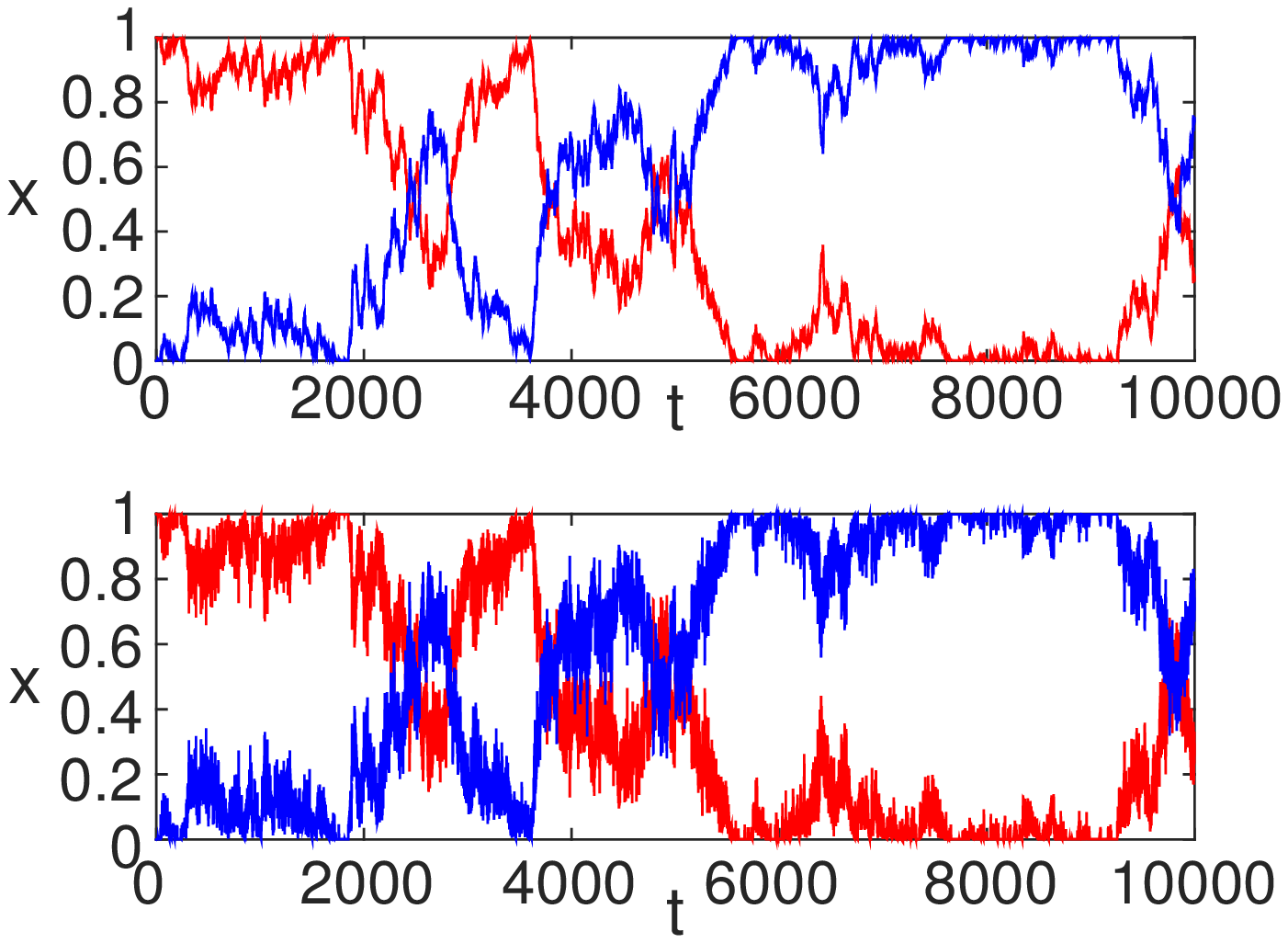}}
  \subfloat
  {\includegraphics[scale=0.3]{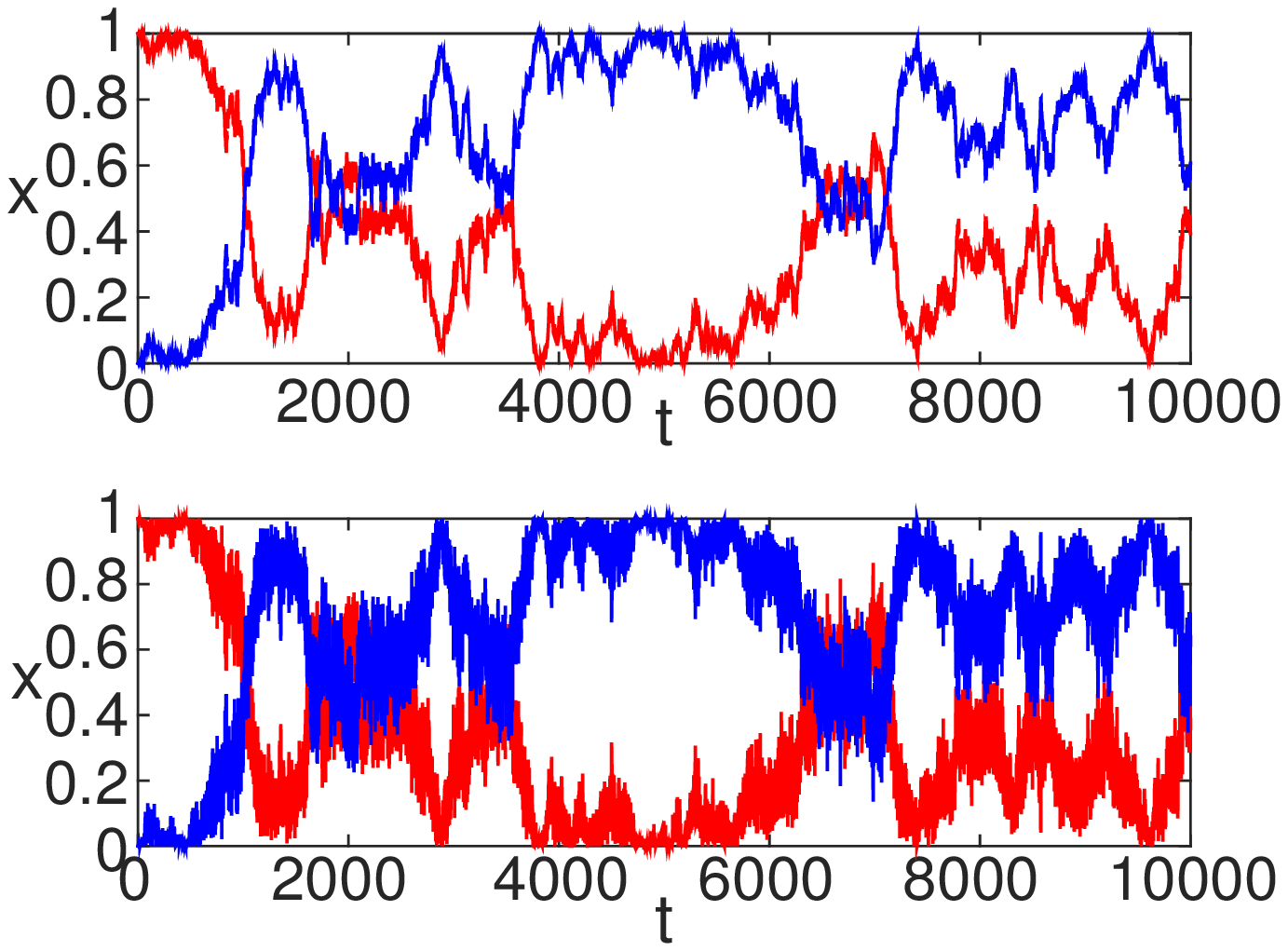}}
  \subfloat
  {\includegraphics[scale=0.3]{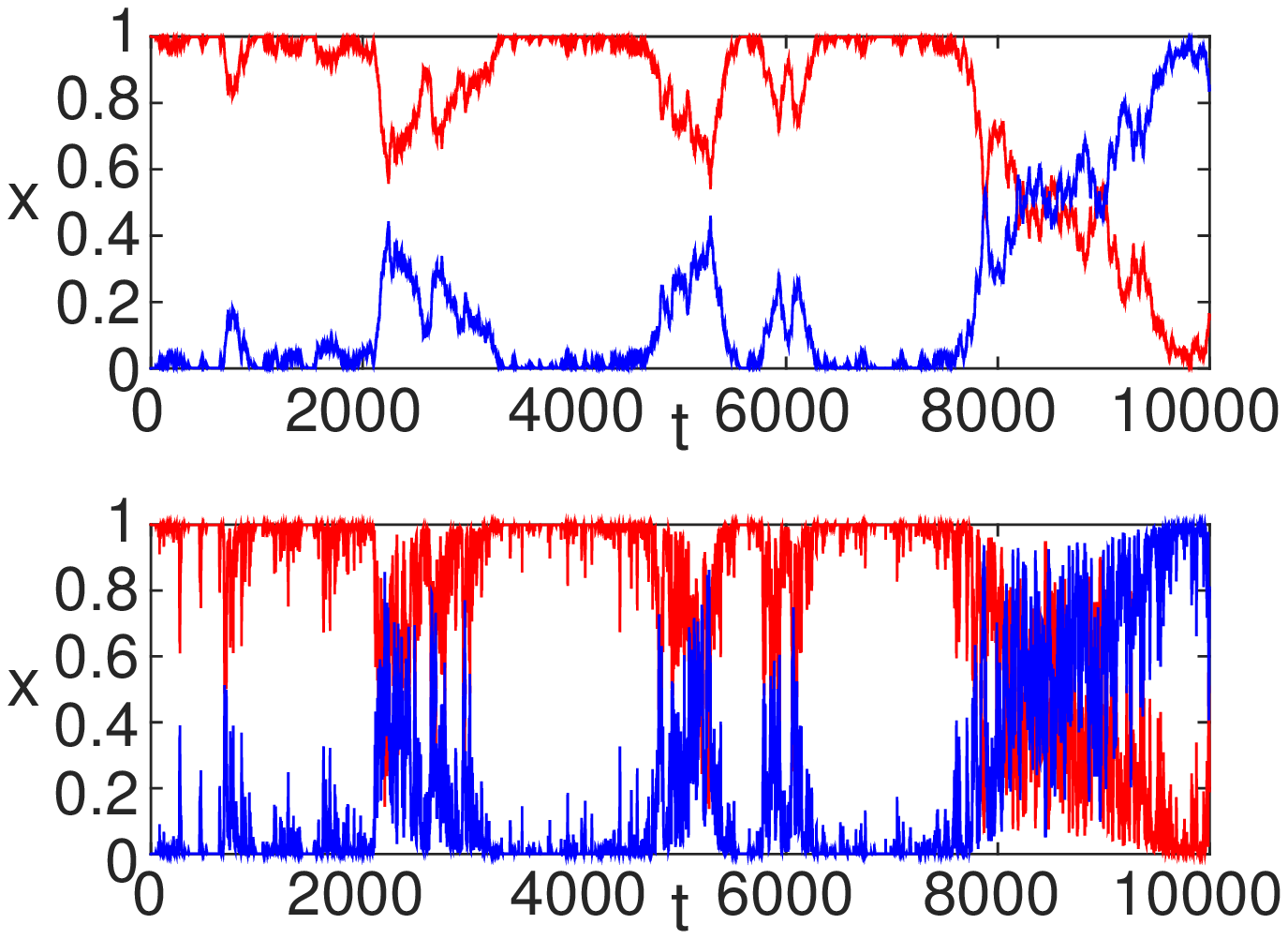}}
  \subfloat
  {\includegraphics[scale=0.3]{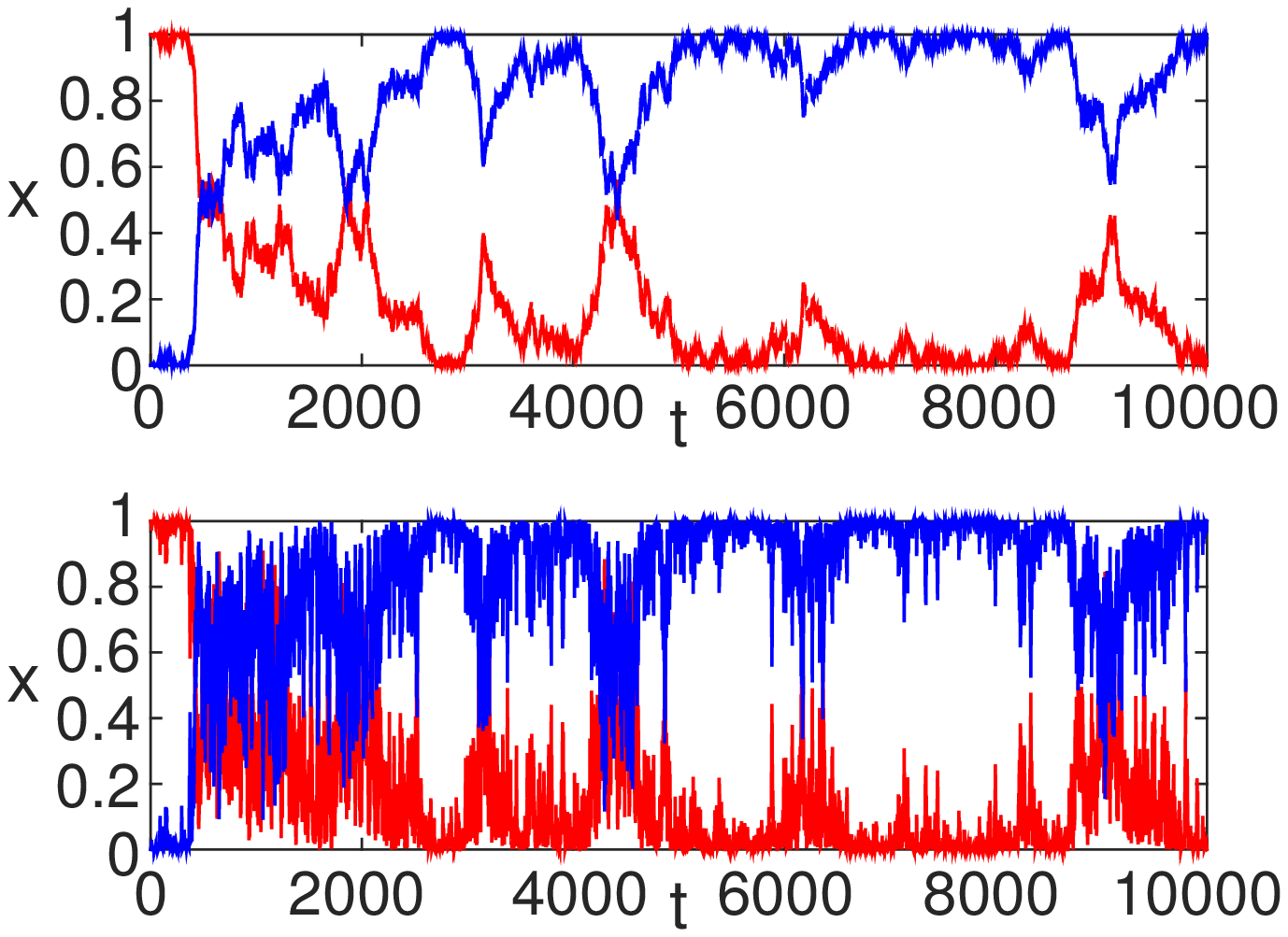}}
  \hspace{5pt}
  \subfloat
  {\includegraphics[scale=0.3]{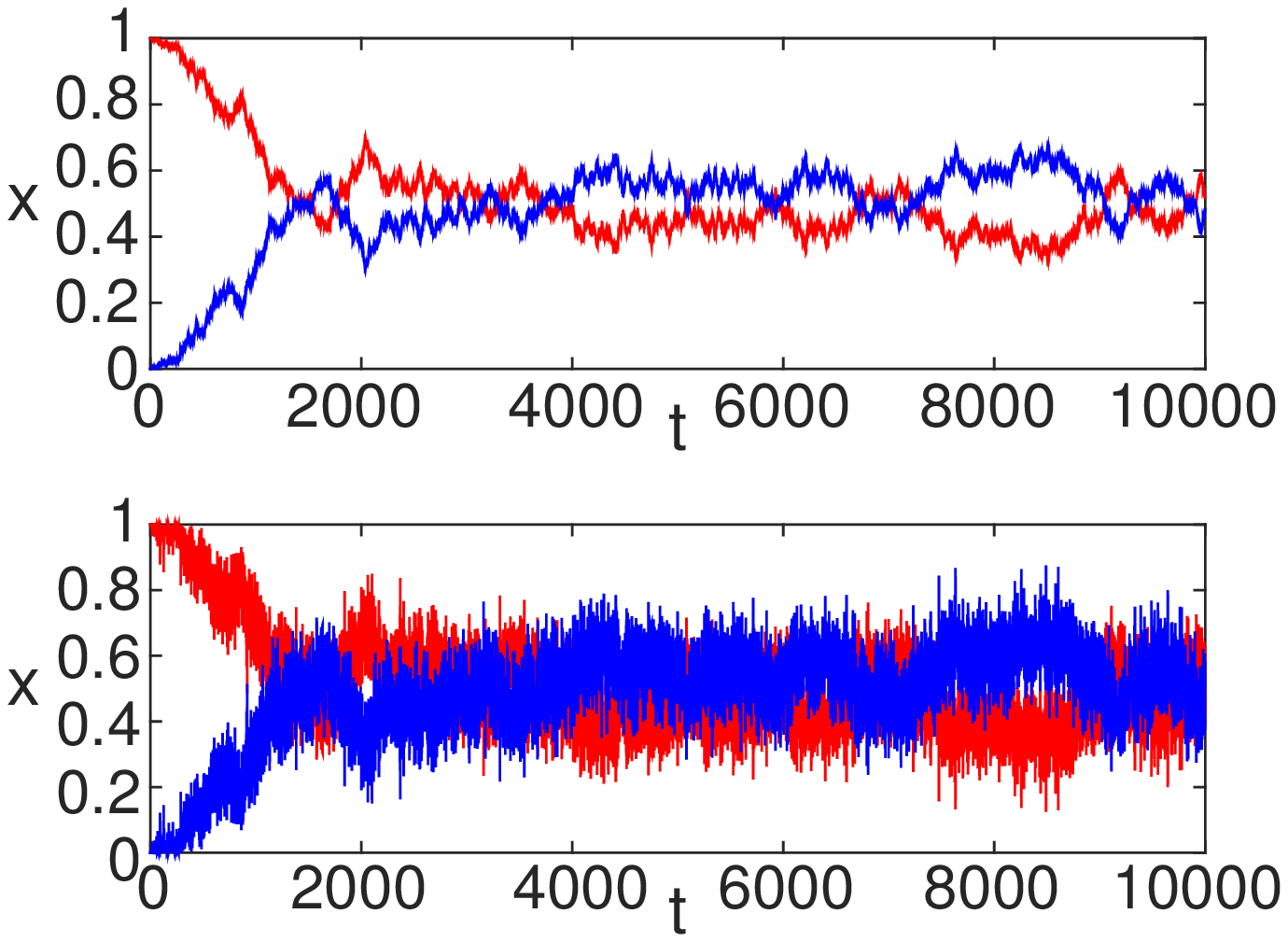}}
  \subfloat
  {\includegraphics[scale=0.3]{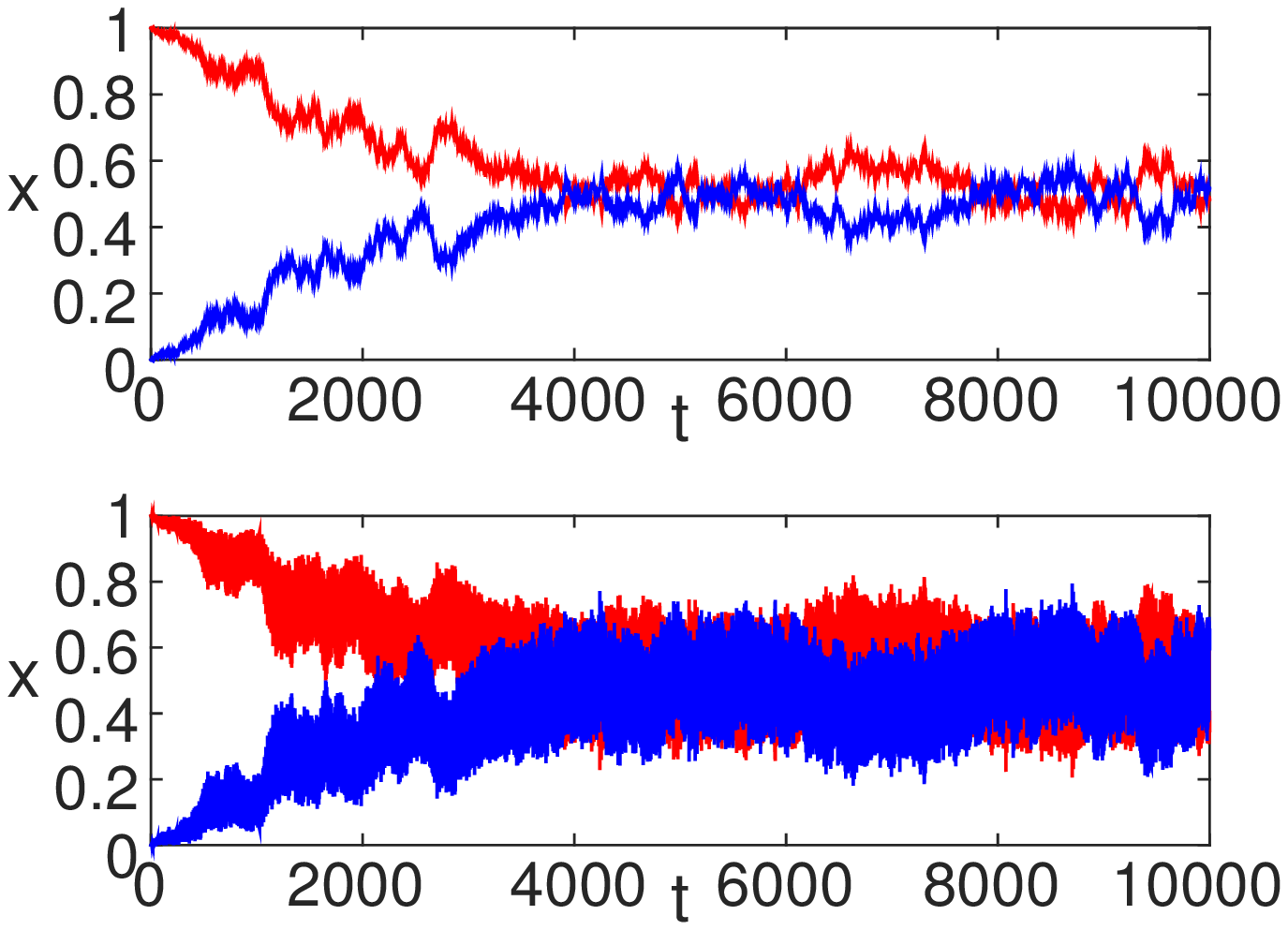}}
  \subfloat
  {\includegraphics[scale=0.3]{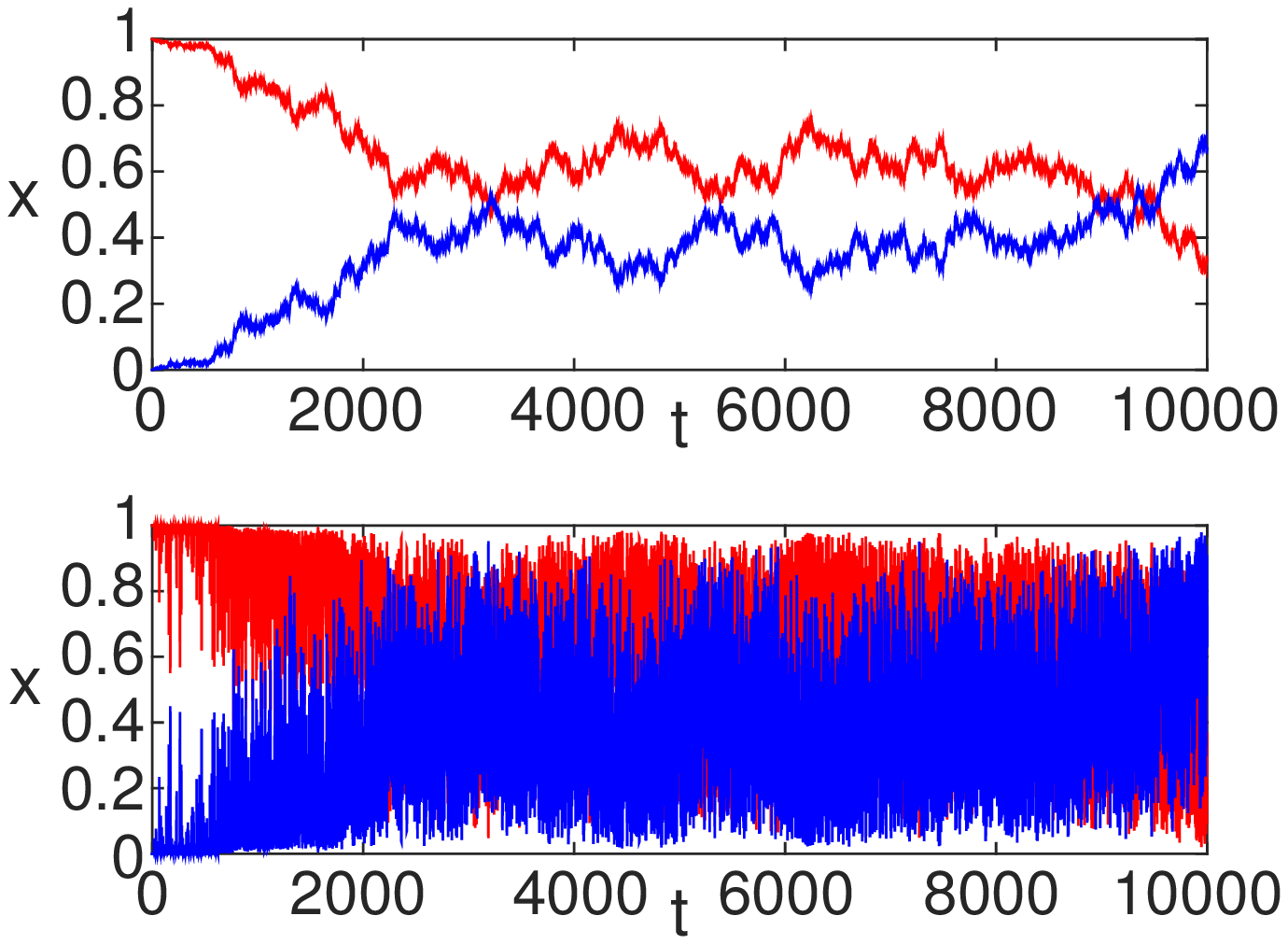}}
  \subfloat
  {\includegraphics[scale=0.3]{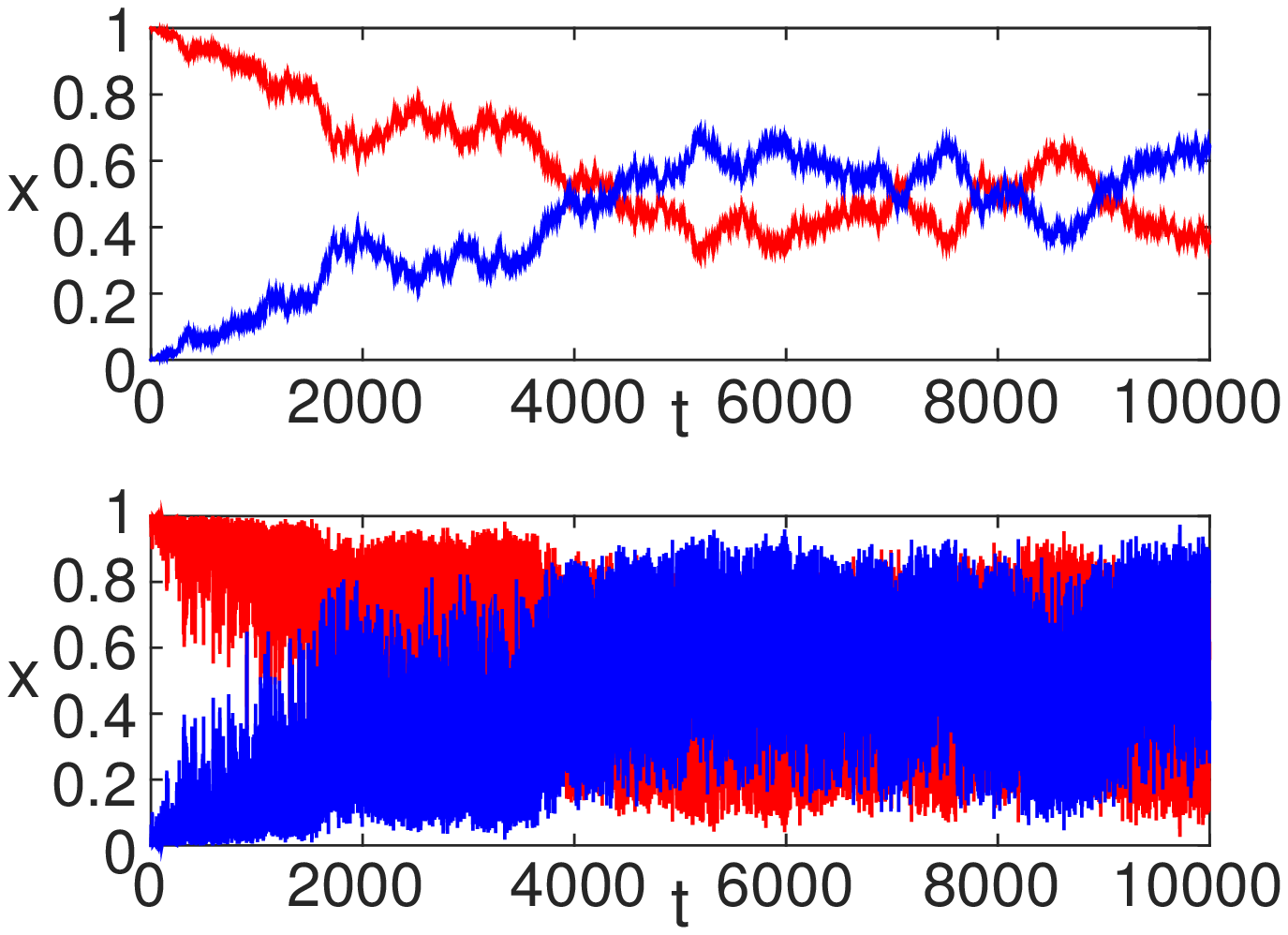}}
  \caption{Comparison of time series of the discrete USM model and of the sHMF limit of it for star-shaped network. For these simulations, the paramaters are $\lambda =0.1$, $V=2$, $L=2$, $T=10^4$ and $q= 0.001$. The upper part of each graph displays the behaviour of the degree $k=1$ nodes and the bottom part of each graph displays the behaviour of the central node. In the first row the value of $N$ is 10 and in the second row, the value of $N$ is 100. The first and third columns display results of the USM and the second and fourth columns display results of the sHMF. In the first two columns $h=0.9$ and in the last two columns $h = 0.1$. At the beginning of the simulation, all agents share the convention to use variant $v=1$.}
  \label{fig:StarNetUSMsHMF}
\end{figure*}

The results are displayed in Fig.~\ref{fig:StarNetUSMsHMF}. Panels (a)--(d) are results for $N=10$ agents and panels (e)--(h) are results for $N=100$ agents. The four left  graphs correspond to $h=0.9$ and the four right graphs correspond to $h=0.1$. We observe that between the $N=10$ and $N=100$ there is a transition from a U-shaped to a bell-shaped distribution. Since the critical value of $q_*$ is derived for the regular network, this existence of a transition should be fairly robust for different topologies. The exact value of $q_*$ is not known for the star-shaped network case, but such a transition is nevertheless expected.

We now discuss the influence of the $h$ parameter. As expected, the behaviour of the central node is noisier than the average of the other nodes. This is a consequence of the time scale difference between the two classes of nodes. If $h$ is reduced, the coupling between the two classes of nodes is weakened and the noise increases. The sHMF reproduces this behaviour and therefore captures the effect of $h$. However, it seems that the sHMF converges with a slower rate towards the stationary distribution. This could be explained by the fact that in the discrete USM, the edges are updated sequentially, whereas in the sHMF they are updated synchronously. The sequential update might converge faster than the synchronous one, as observed in Fig.~\ref{fig:StarNetUSMsHMF}. For large networks, the difference between sequential and synchronous update diminishes and the convergence rates of the two approaches become more similar. Even if the convergence rate of the USM and the sHMF might be different,  the stationary state should nevertheless be similar for both approaches. In order to verify this prediction, we compare the numerical stationary distribution of the USM and of the sHMF in the same conditions as in Fig.~\ref{fig:StarNetUSMsHMF}, computed at $T=4000$. We also compare the results with the predicted mean field approximation corresponding to Eq.~\eqref{eq:HMFRegNet}, where the degree $k$ has to be replaced by the averaged degree $\bar k= 2 + 2/N$ of star-shaped networks. Since the mean field stationary distribution does not depend on $k$, we expect it to be a good approximation if the coupling between the two classes of nodes is strong enough.

\begin{figure*}[t]
  \centering
  \subfloat
  {\includegraphics[scale=0.3]{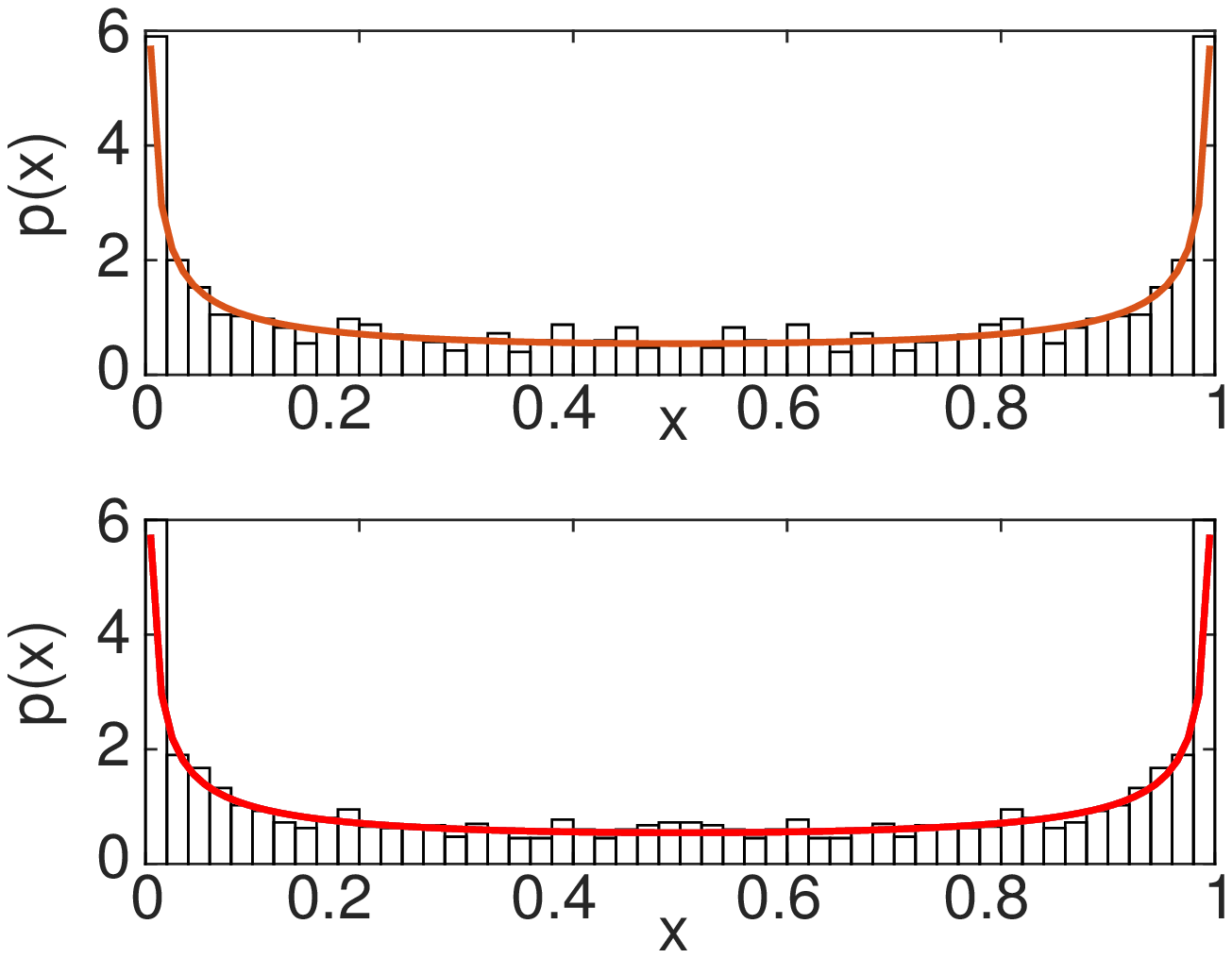}}
  \subfloat
  {\includegraphics[scale=0.3]{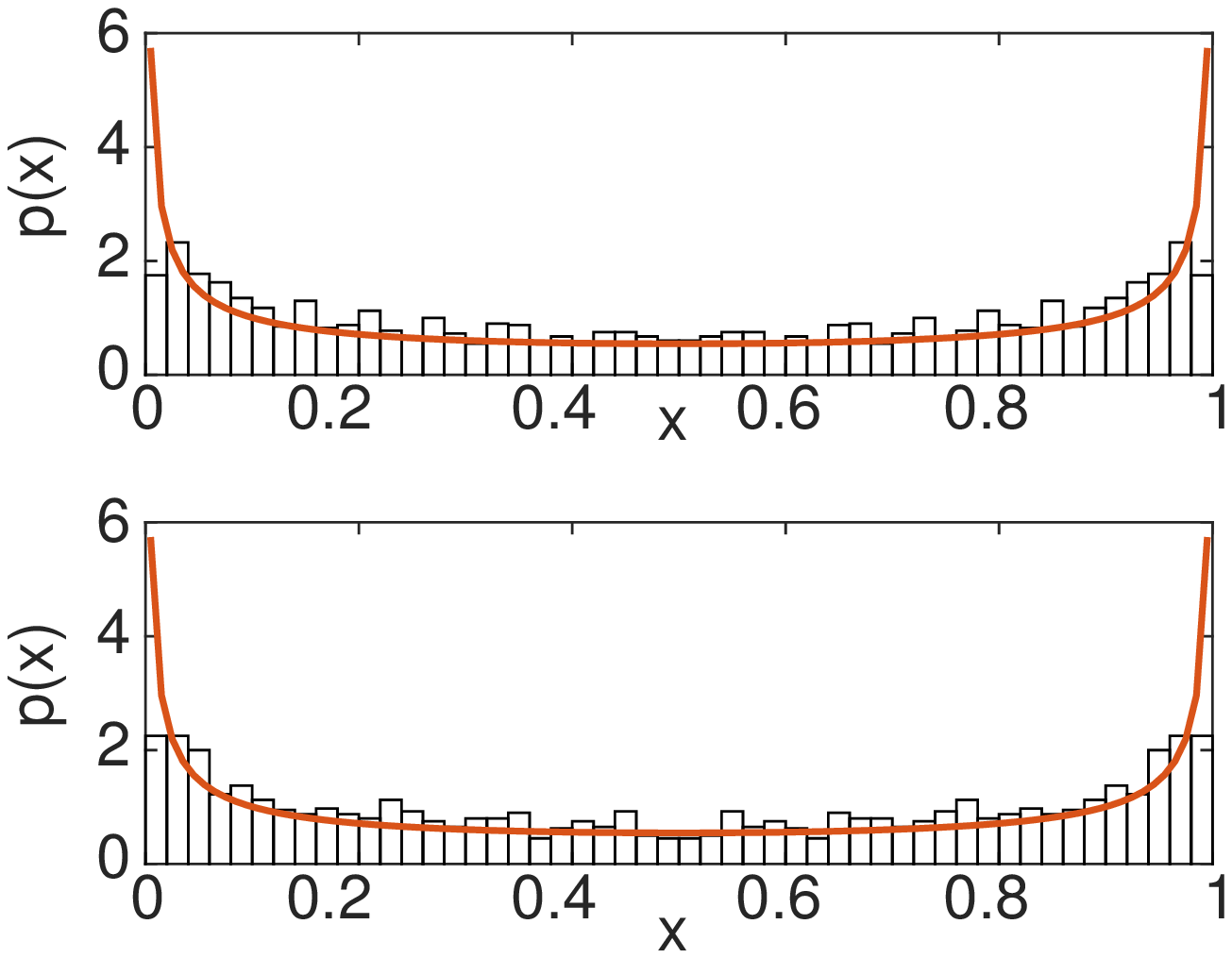}}
  \subfloat
  {\includegraphics[scale=0.3]{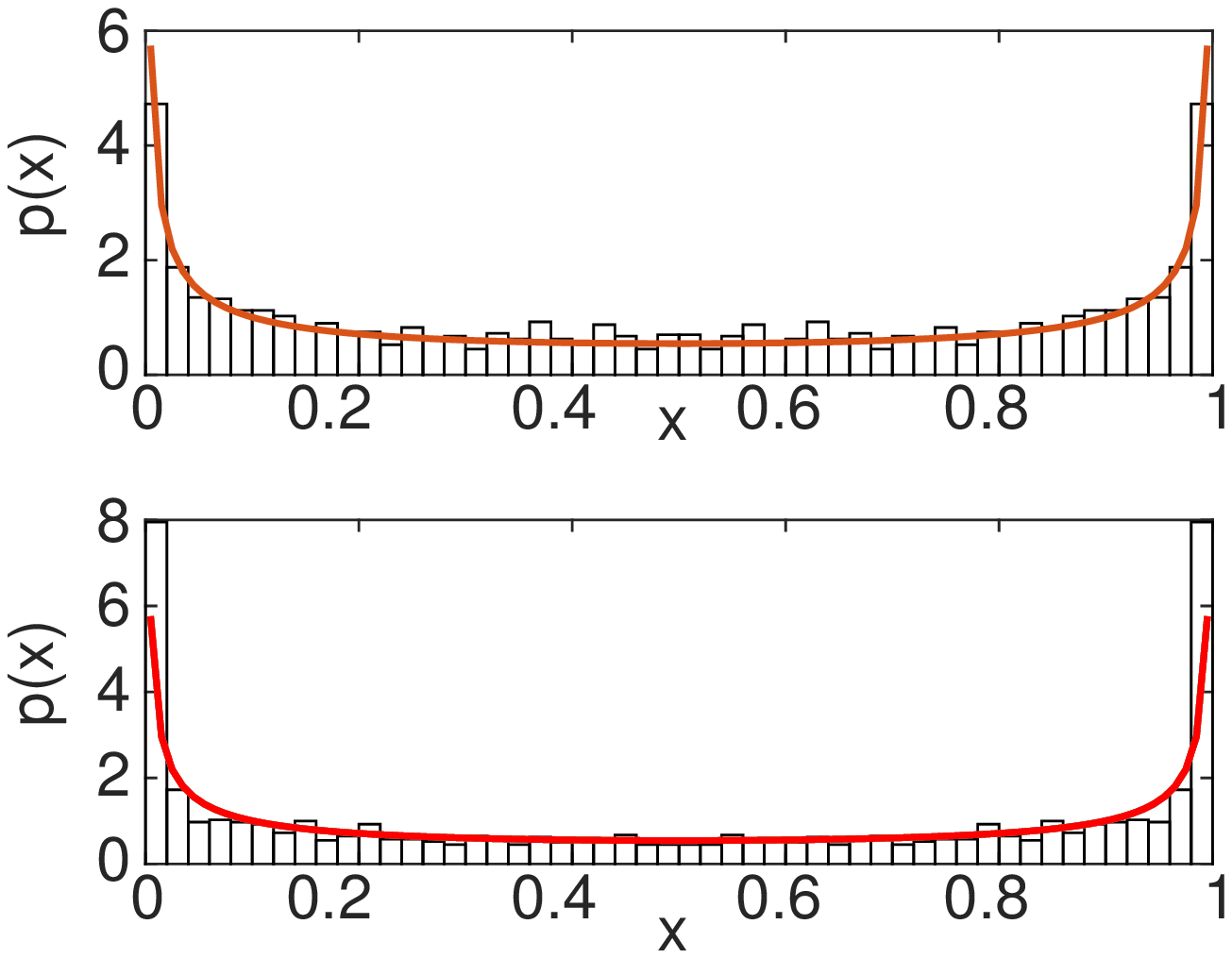}}
  \subfloat
  {\includegraphics[scale=0.3]{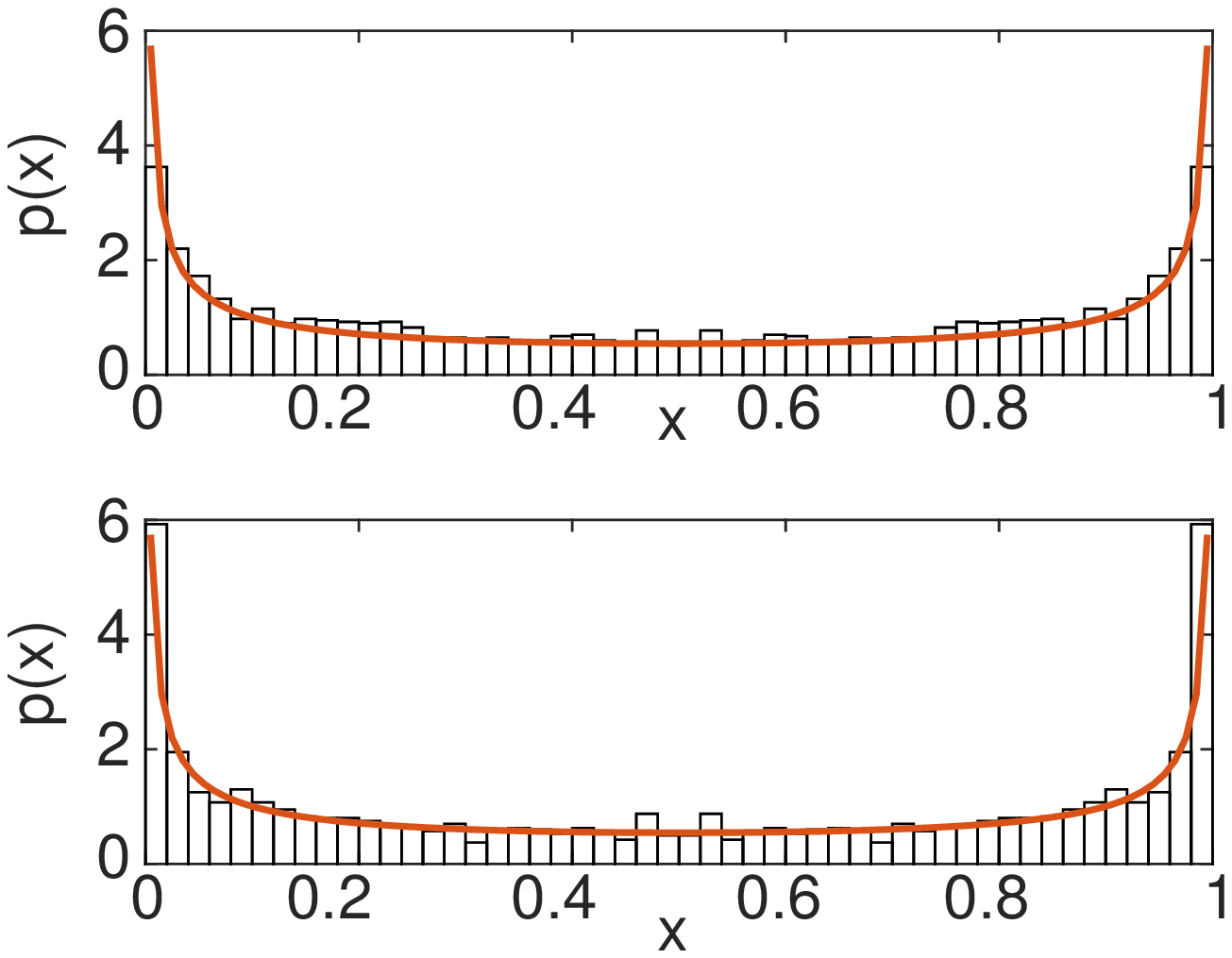}}
  \hspace{5pt}
  \subfloat
  {\includegraphics[scale=0.3]{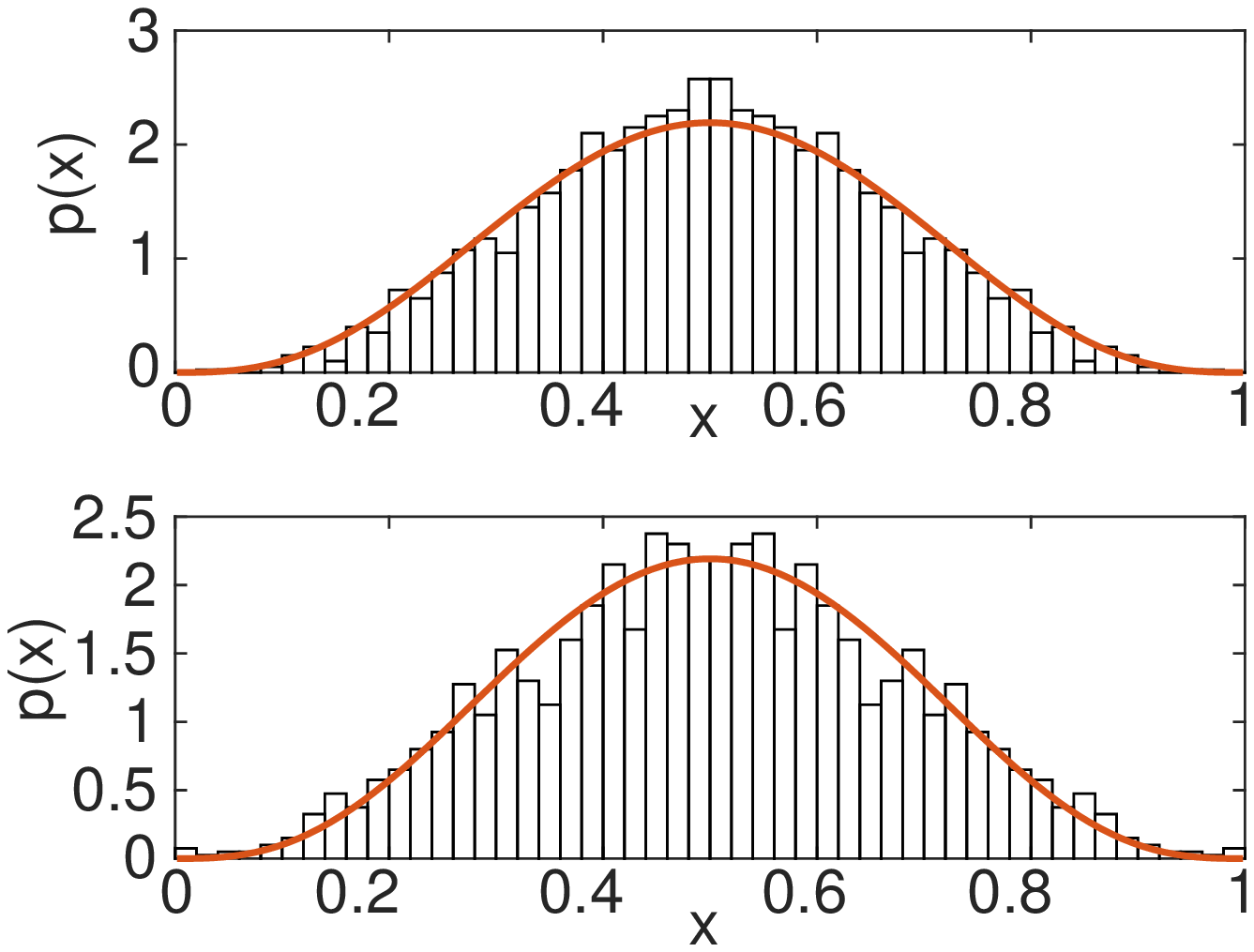}}
  \subfloat
  {\includegraphics[scale=0.3]{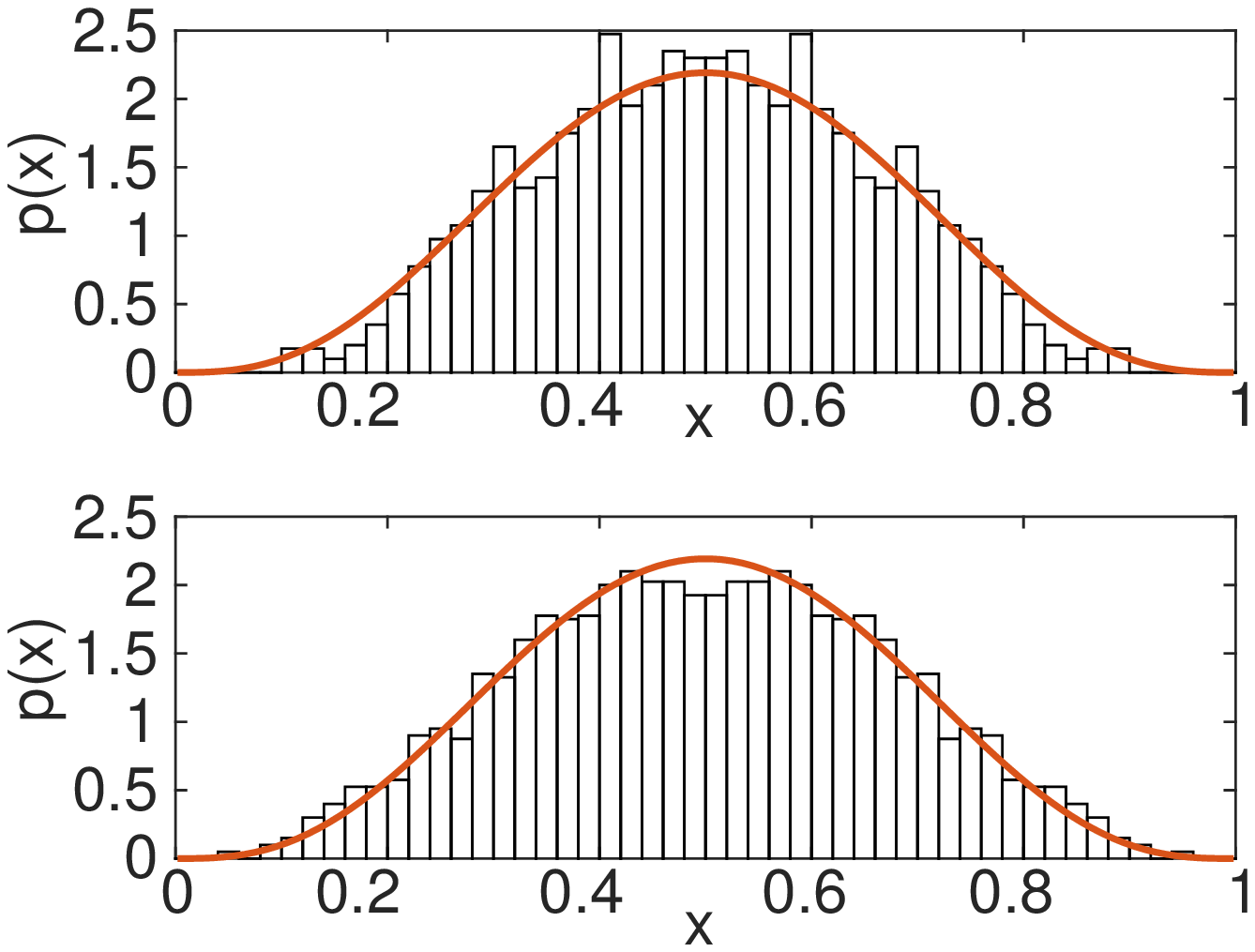}}
  \subfloat
  {\includegraphics[scale=0.3]{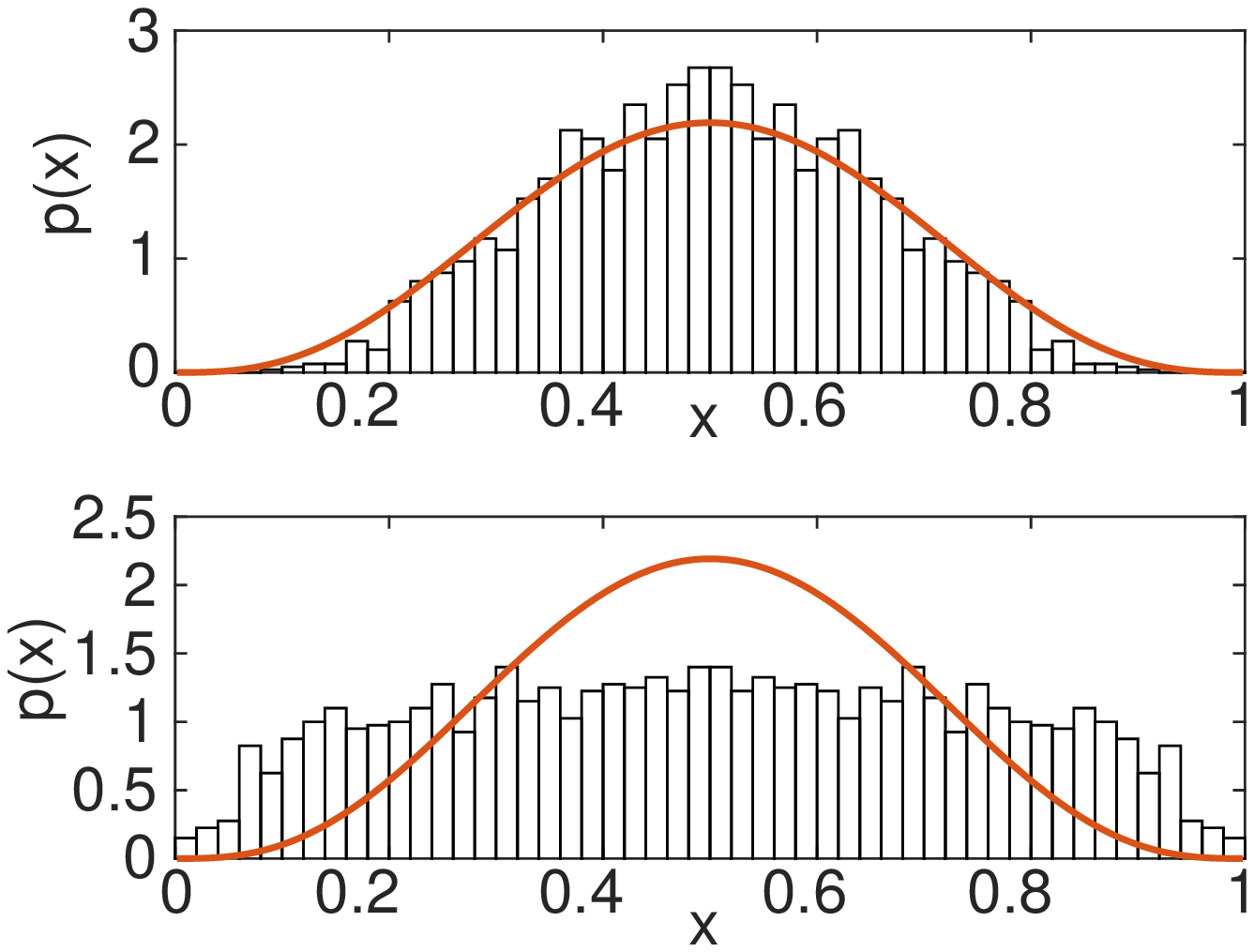}}
  \subfloat
  {\includegraphics[scale=0.3]{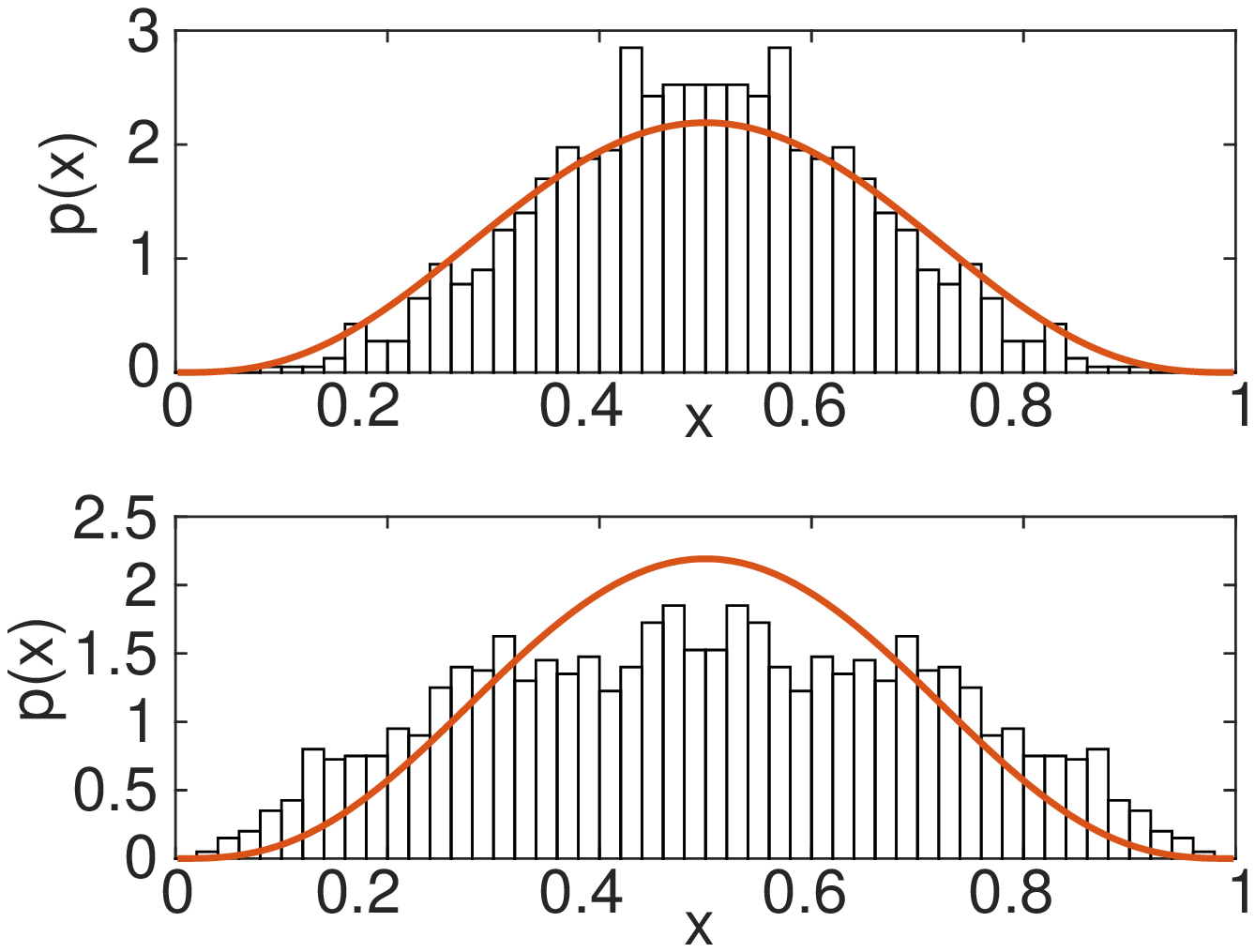}}
  \caption{Comparison between the discrete USM model and its sHMF limit for the star-shaped network. For these simulations, the parameters are $\lambda =0.1$, $V=2$, $L=2$, $T=4\cdot 10^3$ and $q= 0.001$. The upper part of each graph displays the distribution of the degree $k=1$ nodes and the bottom part of each graph displays the distribution of the central node. In the first row the value of $N$ is 10 and in the second row, the value of $N$ is 100. The first and third columns display results of the USM and the second and fourth columns display results of the sHMF. In the first two columns $h=0.9$ and in the last two columns $h = 0.1$. The red line is the solution of the mean field approximation. It helps to see how the star-shaped network differs from the regular network case. The first 2 columns and the last 2 columns have to be compared.}
  \label{fig:histStarNetUSMsHMF}
\end{figure*}

In Fig.~\ref{fig:histStarNetUSMsHMF}, we display the results for the stationary distribution in the same settings as those used in Fig.~\ref{fig:StarNetUSMsHMF}. The sampling is done at the final time $t_{\mathcal G}=T$, $T=4000$. To augment the statistics, we once again considered both $x_1$ and $x_2$ in the histograms, which artificially enforces symmetry of the distributions.

For $h=0.9$, we observe that the distribution of the two degree classes are both in good agreement with the mean field limit. For $N=10$ (Fig.~\ref{fig:histStarNetUSMsHMF}, first row, first two columns), the sHMF is underestimating the behaviour at the boundary of the domain. This might be due to a discretization error effect. In fact, we know that the algorithm used converges strongly (trajectory-wise), but we do not know at which rate. Since this rate can be arbitrarily slow, the results of the sHMF close to the boundary might not be reliable, see App.~\ref{App:NumSim} for details. Apart from this effect, the results of the USM and of the sHMF are in good agreement with the mean field limit (red lines in Fig.~\ref{fig:histStarNetUSMsHMF}). For $h=0.9$ and $N=100$ (Fig.~\ref{fig:histStarNetUSMsHMF}, second row, first two columns), the class of degree $k=1$ nodes follows as expected the mean field limit for both the USM and the sHMF.  The behaviour of the central node is noisier and the comparison is less straightforward. We observe that for the sHMF the effect of the noise manifests itself by an undersampling of the peak of the distribution. This effect is less clear in the USM case, but the results are quite noisy. We can conclude that the sHMF approximation in this case is slightly better that the mean field, which completely neglects the topology of the graph.

For the weaker coupling $h=0.1$, the dynamics becomes more interesting. For $N=10$ (Fig.~\ref{fig:histStarNetUSMsHMF}, first row, last two columns), the agreement between the USM and the sHMF is good and the effect of the stronger noise is mainly seen at the boundaries of the domain, where it is observed that the central nodes spends more time close to the boundary than the average degree $k=1$ node. The discretization problems might explain the undersampling of the sHMF in the boundary regions. Another explanation can be linked with the network reduction itself. In fact, for star-shaped networks, it is not clear whether the approximation should work at all, since the normal approximation fails for the central node. For $N=100$ (Fig.~\ref{fig:histStarNetUSMsHMF}, second row, last two columns), the effect of the noise is much clearer. Since the coupling is fairly weak, the time scale difference between the classes of noise leads to a very noisy behaviour of the central node. As a result, the distribution of the central node flattens, while the behaviour of the degree~$1$ nodes remains close to the mean field. We also observe an oversampling effect of the peak for the degree~$1$ nodes. This can be explained by the fact that in absence of coupling, the behaviour of all the degree~$1$ agents becomes independent. As a consequence of the central limit theorem, the variance of their average behaviour is reduced, explaining the stronger peak observed. The results of the sHMF qualitatively captures the correct behaviour and provide a better prediction than the mean field approximation (red line in Fig.~\ref{fig:histStarNetUSMsHMF}). In this case, the very strong noise entering the dynamics of the central node leads to greater numerical errors, see App.~\ref{App:NumSim} for details. Another explanation of these differences lies in the difference in the variance of the multinomial distribution and of the BISS approximation of it. For instance, in Fig.~\ref{fig:NumExp} it is shown that the variance of the BISS is smaller than the variance of the discrete USM. Since the central node is the only node in his class, the hypothesis based on the central limit theorem, needed to justify the normal approximation, no longer holds. This is a possible explanation of the disagreement of the USM and the sHMF results.  Since the flattening effect is seen in both simulations, we can nevertheless conclude that the sHMF approximation captures the main characteristics of the dynamics of the star-shaped network and in particular the effect of $h$ better than the mean field approximation (red line).

In this section, we have shown that the sHMF approximation is able to capture the dynamics of the USM on different network structures and to reproduce both the trajectories and the stationary distributions of the model. The results of the star-shaped network are less convincing due to numerical problems in the simulation of the sHMF in the presence of strong noise. This is for example the case for $N=100$ and $h=0.1$. Future work will be devoted to finding better algorithms to sample these trajectories.

\section{\label{sec:Conclusion}Conclusion and discussion}
In this paper, we have discussed the USM for language change and its continuous time limits. In order to overcome the parameter restrictions of the FP continuous time limit obtained using the KM expansion, we have proposed a new continuous time limit based on the normal approximation of the multinomial distribution. For two agents, this approximation leads to a weak-noise SDE generalizing the KM expansion solution. We argued that the weak noise should not be neglected for two reasons: (i) the noise is heuristically of the same order of magnitude as the drift term and (ii) the drift term vanishes in long time simulations. The weak-noise limit also captures the influence of the noise of both utterances, whereas the FP limit of \cite{BaxterEtAl06} neglects the influence of the noise of the incoming utterance. 

Using this new continuous time limit, we derived a new stochastic version of the HMF approximation and applied it to regular and star-shaped networks. This approximation allows us to study the dynamics of the system at the level of the network instead of at the level of the agents, which is a great improvement in the analysis of agent-based models in that it provides new analytical tools to characterize the noise-driven phase transition and, therefore, opens the door to new exciting results, since the grouping procedure can be done using different criteria. 

For regular networks, the sHMF formulation turns out to be a Jacobi process described by the WF diffusion SDE. The analysis has shown that the dynamics is controlled by three interdependent parameters: $\lambda k$, $q$ and $r:=\frac{\lambda}{LN}$ and only the last two parameters contribute to the stationary distribution. The $h$ parameter, weighting the self-monitoring and the accommodation process in the USM, does not enter the sHMF approximation. As a result, one can interpret this fact as prestigious agents (large $h$) do not have a particular influence on the dynamics. This is true as long as only the attention parameter is taken into account. If a ``prestigious'' agent influences the weighting of its variants, then the effect can be large. This can for example be modelled by a preference mechanism, see \cite{Michaud16-b}.  For regular networks, we computed the critical value $q_*(r)$ and obtained a phase diagram describing the form of the stationary distribution. Such a distribution is also expected on average for regular graphs, since the sHMF of regular networks can be interpreted as a mean field approximation of any network. Since $r$ is inversely proportional to $N$, the functional dependence $q_*\propto \frac{\lambda}{2LN}$ is the signature of a finite size effect. For instance, the stationary distribution of the averaged population transitions from a U-shaped distribution to a bell-shaped distribution when $N$ increases. In the limit $N\to \infty$, the noise term vanishes and the solution exponentially decays to $x = \frac12$. This case corresponds to the deterministic limit obtained using the KM expansion and only scaling $\lambda = \delta t$.

For star-shaped networks, a case where the sHMF is expected not to be a very good approximation, the sHMF approximation still provides satisfying results, capturing the time scale difference between the central node and the outer nodes. This effect is not captured by the mean field approximation (which corresponds to applying the results from regular networks to star-shaped networks).

In the context of cutural evolution, the interesting regime is when the stationary distribution is U-shaped, which is the signature of the creation of population-wide conventions that can change. In our model, for large populations (for small values of $r$), we have shown that a convention doesn't usually emerge (the stationary distribution is bell-shaped). This is a signature of what is called by Nettle \cite{Nettle99} the \emph{threshold problem}. This problem states that in large populations, it is really difficult to change an established convention. Nettle proposed a solution by using the \emph{Social Impact Theory}. In our case, we can obtain population-wide conventions by increasing $r$ or decreasing $q$, see Fig.~\ref{fig:q}. In other words, one can explain the emergence of new conventions in a large population if the learning rate $\lambda$ is sufficiently large or the if the variability of speech is sufficiently large, that is, if the utterance length $L$ is small. In both cases, the influence of errors is increased. If $q$ is very small, conventions emerge, but they are stable and cultural change is rare. In fact, if $q=0$ the boundaries are \emph{exit} according to Feller classification and conventions are absorbing states.

The Social Impact Theory relies on prestigious agents to explain language change. In the USM, the way the influence of a specific agent is encoded is through the attention parameter $h$. Since this parameter does not enter the mean field equation, our results suggests that an influential agent only has a weak influence on the dynamics. However, if the \emph{prestige} is associated with the variant used by an influential agent, the conclusion changes and this can have a tremendous influence on the dynamics. In this case, the different variants are no longer equivalent and the learning rule has to be adapted to take this into account. Such a variant weighting can be encoded either in the mutation matrix $M$ if the variant is objectively, or functionally, better or through the introduction of a preference mechanism \cite{Michaud16-b}, which allows the agent to adapt their behaviour to the different variants. These modifications have a huge impact on the dynamics of the system and remains to be studied.

In the USM, the influence of the topology can be studied using the sHMF approximation. In this paper, we have provided a proof of principle and the complete analysis of the influence of the network remains to be done. In \cite{Ke08}, the authors have discussed the dynamics of a model of language change in social networks using computer simulations. We believe that our approach can complement and, possibly, explain the results obtained in \cite{Ke08}.

In future work, we will study in detail the influence of the topology of the network using the sHMF and the influence of non-constant $h^{(ij)}$ or asymmetric $M$. We will also study the influence of different extensions of the USM, such as the presence of preferences for a particular variant, the influence of group membership (different behaviour depending of the identity of the interacting agents)... The sHMF only characterizes the stationary distribution of the population-averaged behaviour. Extending this approach to higher moments will complement the knowledge and provide information on the dispersion around the averaged behaviour.

\begin{acknowledgments}
The author has been funded by the Swiss National Science Foundation (SNSF) grant number P2GEP2$\_$159156. The author would like to thanks Richard A. Blythe and Gilles Vilmart for comments and interesting discussions and Edwige Dugas for discussions about the application of this work to linguistics.
\end{acknowledgments}

\appendix
\section{Abbreviations}\label{App:abb}
In this paper, we use a number of abbreviations. In order to ease the reading we list them in Tab.~\ref{tab:Abb}.
\begin{table}[h]
\caption{List of abbreviations}
\label{tab:Abb}
\begin{tabular}{ll}
\hline\hline
\multicolumn{2}{c}{\bf General abbreviations}\\
\hline
USM&Utterance selection model\\
(s)HMF&(stochastic) heterogeneous mean field\\
KM&Kramers-Moyal\\
FP&Fokker-Planck\\
SDE&Stochastic differential equation\\
WF& Wright-Fisher\\\hline
\multicolumn{2}{c}{\bf Numerical methods}\\\hline
EM&Euler-Maruyama\\
IL&Internal limiter\\
EL&External limiter\\
MS& Moro-Schurz\\
BIM& Balanced implicit method\\
BISS&Backward implicit split step\\
\hline\hline
\end{tabular}
\end{table}
\section{\label{App:USM}Continuous time limit of the USM using Kramers-Moyal expansion}

In this appendix, we provide the derivation of continuous time limits of the USM using the Kramers-Moyal (KM) expansion \cite{Risken84}, similarly to what has been done in \cite{BaxterEtAl06}.
This method provides a Fokker-Planck (FP) equation for the probability distribution $p({\vec x}{^{(i)}},t; \vec X^{(-i)})$ to find agent $i$ with an idiolect $\vec x^{(i)}$ at time $t$, knowing the state of the rest of the population $\vec X^{(-i)}$. The exponent $(-i)$ means: all agents except agent $i$. This is a notation borrowed from game theory. The time $t$ has to be measured in $t_{\rm int}$ units here.

The KM expansion of a stochastic process $\vec x^{(i)}$ is given by
\begin{equation}
\begin{aligned}
\frac{\partial p}{\partial t} =&- \sum_{v=1}^{V-1} \frac{\partial}{\partial x^{(i)}_{v}}\{\beta_v(\vec x^{(i)})p\}\\
&+ \frac{1}{2}\sum_{v=1}^{V-1}\sum_{w=1}^{V-1} \frac{\partial^2}{\partial x^{(i)}_{v}\partial x^{(i)}_{w}}\{\beta_{vw}(\vec x^{(i)})p\}\\&+\dots
\end{aligned}
\end{equation}
where the jump moments are defined as
\begin{eqnarray}
\beta_v(\vec x^{(i)}) &=& \lim_{\delta t \to 0}\frac{\langle\delta x^{(i)}_v(t)\rangle}{\delta t},\\ \beta_{vw}(\vec x^{(i)}) &=&\lim_{\delta t \to 0}\frac{\langle\delta x^{(i)}_v(t)\delta x^{(i)}_w(t)\rangle}{\delta t}.
\end{eqnarray}
Here, the average is taken over utterance production and over edges of the graph connected to agent $i$, which are the two sources of randomness in the model.

In order to simplify a bit the presentation, we assume that the off-diagonal terms of the matrix $M$ are of order $(\delta t)^{1/2}$ or smaller. If this is the case, then one can write the condition
\begin{equation}\label{eq:Ass-off}
O(\|M-I\|_\infty) = O((\delta t)^{1/2}).
\end{equation}
This assumption has been used in \cite{BaxterEtAl06} and we only do it in this appendix. 
We also introduce the notation $\vec x' = M\vec x$ for convenience.

Under assumption \eqref{eq:Ass-off} one can collect all the terms that depend on the off-diagonal term of $M$ in $O(\|M-I\|_\infty)$. We can then write the first two jump moments as
\begin{equation}\label{eq:jm1}
\langle\delta x^{(i)}_v\rangle = \sum_{j\neq i}G^{(ij)}\lambda \Big[(1-h)( x'^{(i)}_v-x^{(i)}_v) + h(x'^{(j)}_v-x^{(i)}_v)\Big],
\end{equation}
and
\begin{equation}\begin{aligned}\label{eq:jm2}
\langle\delta x^{(i)}_v\delta x^{(i)}_w\rangle\! =&\! \sum_{j\neq i}\!G^{(ij)}\lambda^2 \Bigg[\frac{(1-h)^2}{L} x^{(i)}_v(\delta_{vw} - x^{(i)}_w)\\& +\! \frac{h^2}{L}x^{(j)}_v(\delta_{vw} -  x^{(j)}_w)\\
&+ h(1-h)(x^{(j)}_w - x^{(i)}_w)(x^{(j)}_v - x^{(i)}_v)\\& + O(\|M-I\|_\infty)
\Bigg].
\end{aligned}\end{equation}

Equation \eqref{eq:jm2} has been computed for the definition \eqref{eq:utterBS} of the utterances. The expression for the definition \eqref{eq:utterSB} differs from \eqref{eq:jm2} and can be computed easily.  

In order to obtain a Fokker-Planck equation, scaling assumptions have to be made to ensure that \eqref{eq:jm1} and \eqref{eq:jm2} are both of order $\delta t$ and that higher order jump moments are of higher order. The scaling chosen in \cite{BaxterEtAl06} is given by
\begin{subequations}
\label{eq:scalingUSM}
\begin{eqnarray}
\lambda &=& (\delta t)^{1/2},\\
M_{vw} &=& \bar M_{vw}(\delta t)^{1/2},\quad \text{for }v\neq w,\label{eq:scalingUSM-m}\\
h &=& \bar h(\delta t)^{1/2}.
\end{eqnarray}
\end{subequations}
Eq.~\eqref{eq:scalingUSM-m} is equivalent to assumption \eqref{eq:Ass-off}.
This scaling is the only one compatible with the KM expansion leading to a FP equation with non-vanishing diffusion, given the constraints on the parameters. In particular, if the constraint that $L$ is an integer is relaxed,  another scaling would work. It is given by
\begin{subequations}
\label{eq:scalingL}
\begin{eqnarray}
\lambda&=&\delta t,\\
L&=&\bar L \delta t.
\end{eqnarray}
\end{subequations}
The scaling of $L$ means that the number of tokens in an utterance tends to $0$. Since $L\geq 1$, this is not possible.

The USM scaling is problematic since it requires to scale the $h$ parameter and the off-diagonal terms of $M$, limiting this continuous time limit to a small part of the parameter space. The assumption that off-diagonal terms of $M$ are small corresponds to a small probability of innovation and is not really problematic. The restriction on $h$ is much stronger, since it requires the accommodation process to be negligible with respect to the self-monitoring process, which is usually not justified.

The second scaling is not satisfying either since it requires to scale an integer quantity, namely $L$. Therefore, none of these approaches give a satisfying FP equation. 

If one does not want to scale either $h$ or $L$, the only possible scaling left is to scale $\lambda \propto \delta t$. In this case, the KM expansion is truncated after the first term and there is no diffusion term. In other words, the continuous time limit is deterministic.

The KM expansion, therefore, predicts that the behaviour of a single agent on the $t_{\rm int}$ time scale is deterministic, unless the attention parameter $h$ and the off-diagonal terms of $M$ are small, in which case, we obtain a diffusive dynamics. 


\section{\label{App:WF}USM and the Wright-Fisher process}
In this appendix, we present the Wright-Fisher (WF) stochastic process, also called Jacobi process, and connect it to the USM. We then discuss the different available choices of chosing a noise form in the resulting stochastic differential equation (SDE).
\subsection{Definition of the Wright-Fisher process}
The WF models of population genetics \cite{Wright31,Fisher30} models the biological transmission of alleles of genes between generations of a population. This model give rise to a stochastic process described by the SDE
\begin{equation}\label{eq:WFproc}
d\vec x_t = -\lambda(\vec x_t- \vec b) + c({\rm diag}(\vec x_t)-\vec x_t\vec x_t^T)^{1/2} d\vec W_t,
\end{equation}
where $\lambda>0$, $\vec b\in \mathds P_V$, $c$ is a positive constant and the square root of the matrix has to be taken in the Cholesky sense. Finally $d\vec W_t$ is a $d$ dimensional white noise. $d$ is not necessarly equal to the dimension $V$ of $\vec x_t$, since the Cholesky square root is not necessarily a square matrix. Note that one only needs to consider the first $V-1$ components of $\vec x_t$, since the last one can be recovered using the conservation of probability.
\begin{defi}[Square root in the Cholesky sense]\label{def:SCholesky} A matrix $D \in \mathds R^{m\times n}$ is said to be a \emph{square root in the Cholesky sense} of a matrix $A \in \mathds R^{m\times m}$ if
\[
DD^T = A.
\]
The square root in the Cholesky sense is not uniquely defined and not necessarily a square matrix, see \cite{StroockVaradhan07} for details.
\end{defi}

The WF process \eqref{eq:WFproc} satisfies a sum to unit constraint and a non-negativity constraint. In \cite{BakosiRistorcelli14} it is shown that there are only a few stochastic processes that satisfy such a conservation law. The WF process naturally arises from discrete processes when  a characterized by a multinomial sampling process. The is the case in the original discrete WF model as well as in the USM. For instance, the matrix ${\rm diag}(\vec x_t)-\vec x_t\vec x_t^T$ corresponds to the covariance matrix of the normalised multinomial sampling process.

The WF stochastic process is sometimes called the Jacobi process by mathematician and economists \cite{GourierouxValery04,GourierouxJasiak06,kuznetsov2004solvable,KarlinTaylor81}, because the infinitesimal generator of this process, obtained as the eigenfunctions of the backward Kolmogorov equation, are Jacobi polynomials.

We now discuss the form of the matrix $D(\vec x)$, which is the Cholesky square root of the matrix
\begin{equation}
 A(\vec x):= {\rm diag}(\vec x_t)-\vec x_t\vec x_t^T
\end{equation}
 in the Cholesky sense. This matrix is needed to complete the formulation of the WF process \eqref{eq:WFproc} and is also used in the normal approximation \eqref{eq:Normalw} assumed in the derivation of continuous time limits of the USM. We start with the special case of $V=2$ and discuss then the general case.



\subsection{Form of $D$ when $V=2$}
In the case $V = 2$, a vector $\vec x \in \mathds P_2$ is such that $x_2 = 1-x_1$ and the matrix $A(\vec x)$ takes the simple form
\begin{equation}\label{eq:Cov}
A(\vec x) = x_1(1-x_1)\begin{bmatrix}1&-1\\-1&1\end{bmatrix}.
\end{equation}
This matrix has many Cholesky square roots. We list three of them here.
\begin{subequations}\label{eq:CholSQR}
\begin{eqnarray}
D_1(\vec x) &:=& \frac{1}{\sqrt{2}}\sqrt{x_1(1-x_1)}\begin{bmatrix}1&-1\\-1&1\end{bmatrix},\label{eq:Ds.1}\\
D_2(\vec x) &:=& \sqrt{x_1(1-x_1)}\begin{bmatrix}1\\-1\end{bmatrix},\label{eq:Ds.2}\\
D_3(\vec x) &:=& \begin{bmatrix}(1-x_1)\sqrt{x_1}&-x_1\sqrt{1-x_1}\\-(1-x_1)\sqrt{x_1}&x_1\sqrt{1-x_1}\end{bmatrix}.\label{eq:Ds.3}
\end{eqnarray}
\end{subequations}
It is straightforward to check that these matrices are Cholesky square roots of \eqref{eq:Cov}. The matrix $D_1(\vec x)$ is also a square root of $A(\vec x)$ in the sense that $D_1(\vec x)D_1(\vec x) = A(\vec x)$. For simplicity, the matrix $D_2(\vec x)$ is usually chosen. 

\subsection{General case}
If $V>2$, than one can generalize the choices \eqref{eq:Ds.2} and \eqref{eq:Ds.3}, but the choice \eqref{eq:Ds.1} is more difficult to generalize. 

The choice \eqref{eq:Ds.1} corresponds to Cholesky square root that is also a square root in the sense that $D_1^2 = A$. Finding matrix square roots is not an easy task and, therefore, this choice is difficult to generalize. 

The choice \eqref{eq:Ds.2} takes into account the possible reaction channels and consider one noise for each. For example, if $V=3$ then there are three mutation channels: $x_1 \leftrightarrow x_2$, $x_1 \leftrightarrow x_3$ and $x_2 \leftrightarrow x_3$ and a possible Cholesky square root is given by
\begin{equation}\label{eq:D.WF}
D(\vec x):= \begin{bmatrix}
\sqrt{x_1x_2}&\sqrt{x_1x_3}&0\\
-\sqrt{x_2x_1}&0&\sqrt{x_2x_3}\\
0&-\sqrt{x_3x_1}&-\sqrt{x_3x_2}
\end{bmatrix}.
\end{equation}
This can be generalized to an arbitrary number of variants. The dimension of this matrix is $V\times {{V}\choose{2}}$, where ${{V}\choose{2}}$ is the binomial coefficient. This is for example the formulation used in \cite{DangerfieldKayMacNamaraBurrage12}.

The generalization of Eq.~\eqref{eq:Ds.3} is given by 
\begin{equation}\label{eq:D.Jacobi}
\{D(\vec x)\}_{vw} :=(\delta_{vw}-x_v)\sqrt{x_w}.
\end{equation}
This choice is associated with the multivariate Jacobi process, see \cite{GourierouxJasiak06}.

All these choices are equivalent. In the context of SDEs, they correspond to different trajectories of the same Wiener process, see for example \cite{StroockVaradhan07}.

\subsection{The USM and the WF process}
We now detail a case in which the USM is related to the WF process. We consider a network of $N=2$ agents using $V=2$ variants.
Given a mutation matrix $M$ of the form
\begin{equation}\label{eq:M2}
M:=\begin{bmatrix}1-m_2&m_1\\m_2&1-m_1\end{bmatrix},
\end{equation}
from Eq.~\eqref{eq:CTLSDE} we obtain the following equations
\begin{subequations}
\label{eq:CTLSDE3}
\begin{equation}
d\vec x^{(i)}=G^{(i)} \Bigg[
\left((M-I)\vec x^{(i)} \right)dt + \left(\frac{1}{\sqrt{L}}D(M\vec x^{(i)})d\vec \xi^{(i)} \right)\Bigg],
\end{equation}
or
\begin{equation}\label{eq:CTLSDE3.2}
d\vec x^{(i)}=G^{(i)} \Bigg[
\left((M-I)\vec x^{(i)} \right)dt + \left(\frac{1}{\sqrt{L}}MD(\vec x^{(i)})d\vec \xi^{(i)} \right)\Bigg],
\end{equation}
\end{subequations}
where the matrix $D(\vec x)$ is given by Eq.~\eqref{eq:Ds.2}. We then have
\begin{subequations}
\begin{eqnarray}
D(M\vec x^{(i)})&=&\sqrt{x'^{(i)}_1(1-x'^{(i)}_1)}\begin{bmatrix}1\\-1\end{bmatrix},\\
\label{eq:2V-D}MD(\vec x^{(i)})&=&(1-m_1-m_2)\sqrt{x^{(i)}_1(1-x^{(i)}_1)}\begin{bmatrix}1\\-1\end{bmatrix},\ \qquad
\end{eqnarray}
\end{subequations}
where $x'_1$ is the first component of $\vec x' = M\vec x$.  

As stated in section \ref{sec:USM} the components of $\vec x^{(i)}\in \mathds P_2$ are not independent and it is sufficient to only consider the evolution of the first components. We obtain
\begin{subequations}
\label{eq:WF}
\begin{eqnarray}
dx^{(i)}_1 &=&-\gamma(x^{(i)}_1-\mu)dt + \sigma_{\rm sb}\sqrt{x^{(i)}_1(1-x^{(i)}_1)}dW_t^{(i)},\label{eq:WF.1}\qquad\\
dx^{(i)}_1 &=&-\gamma(x^{(i)}_1-\mu)dt + \sigma_{\rm bs}\sqrt{x'^{(i)}_1(1-x'^{(i)}_1)}dW_t^{(i)},\ \qquad\label{eq:WF.2}
\end{eqnarray}
\end{subequations}
where 
\begin{equation}
\begin{aligned}
\gamma&:= -G^{(i)}(m_1+m_2);\\
\mu&:=\frac{m_2}{m_1 +m_2};\\
\sigma_{\rm sb}&:=\frac{\sqrt{dt}G^{(i)}}{\sqrt{L}}(1-m_1-m_2);\\
\sigma_{\rm bs}&:=\frac{\sqrt{dt}G^{(i)}}{\sqrt{L}}.
\end{aligned}
\end{equation}

We now discuss the influence of the ordering of sampling and biasing on this weak-noise SDE. Since Eq.~\eqref{eq:WF} has to satisfy the constraint that $x_1 \in [0,1]$, the SDE has to satisfy a number of properties discussed in \cite{BakosiRistorcelli14}. One of these properties is that the noise coefficient has to vanish at the boundaries of the interval, that is at $x_1 = 1$ and $x_1=0$. The property is satisfied by Eq.~\eqref{eq:WF.1}, but not by Eq.~\eqref{eq:WF.2}. One can, therefore, conclude that Eq.~\eqref{eq:WF.2} is ill-posed, since it does not conserve the probability. The well-posedness of Eq.~\eqref{eq:WF.1} then follows from the Yamada-Watanabe theorem \cite{YamadaWatanabe71}. This theorem has to be used because the noise coefficient is not a Lipschitz continuous function. Recall that a Lipschitz continuous function $f$ satisfies
\begin{equation}\label{eq:Lipschitz}
\|f(x)-f(y)\|_2 \leq C_L |x-y|,\quad \forall x,y \in \mathcal D(f),
\end{equation}
where $C_L$ is the Lipschitz constant and $\mathcal D(f)$ is the domain of $f$. This non-Lipschitz aspect of the noise coefficient in Eq.~\eqref{eq:WF} is at the origin of numerical difficulties, see App.~\ref{App:NumSim}.

Under the normal approximation, the order of the sampling and biasing processes matters. Sampling first and then biasing is the only one that leads to a well-posed SDE. This order is also more natural in a linguistic framework, it corresponds to first sampling for the belief distribution $\vec x$ and then modifying the output as a result of passing through the articulatory-auditory channel. The other ordering corresponds to modifying the belief distribution $\vec x$ and then sampling from the biased distribution $\vec x' = M\vec x$ without error. The origin of errors is more difficult to justify in this case. The most natural ordering is then also the mathematically preferred. Note that in the USM and in the Dirichlet approximation, both orders are possible and the restriction obtained here is intrinsically connected with the normal approximation and its unbounded nature, see Sec.~\ref{sec:approx}. 
The discussion about the well-posedness of the equation has been done for two variants. Using the results of \cite{BakosiRistorcelli14}, one can generalize the results to an arbitrary number of variants and we arrive at the same conclusion that the only ordering leading to a well-posed equation is sampling first and then biasing.

Another example in which the USM is linked with the WF model is given by the sHMF approximation for regular graphs given in Sec.~\ref{ssec:RegNet}.

\section{\label{App:NumSim}Numerical algorithms}
In this appendix, we discuss the possible numerical strategies to solve the Wright-Fisher SDE occuring as the sHMF of the regular network and how to extend the results to the general sHMF equation. We consider the SDE
\begin{equation}\label{eq:WFA}
dx_t = -\gamma(x_t-\mu)dt + \sigma \sqrt{x_t(1-x_t)}dW_t,
\end{equation}
where $x_t$ is a realization of the stochastic process and $dW_t$ a white noise. Eq.~\eqref{eq:HMFRegNet} is of this form. This equation can be shown to be well-posed on $[0,1]$ using a result of Yamada and Watanabe \cite{YamadaWatanabe71}. One difficulty that arises with this kind of SDE is linked with the non-Lipschitz aspect of the multiplicative noise. Most of the usual proof of convergence rely on a Lipschitz condition \eqref{eq:Lipschitz}. 

In order to accurately capture the trajectory of the stochastic process, one needs a strongly convergent numerical method, see \cite{KloedenPlaten92} for details about the types of convergence. There is a weaker notion of convergence, known as weak convergence, that only requires convergence on avearage and not trajectory-wise. Obtaining weakly convergent methods is usually much easier than obtaining strongly convergent methods.

In the rest of this appendix, we discuss the performance of different numerical methods for integrating Eq.~\eqref{eq:WFA}, we then obtain a numerical method to integrate Eq.~\eqref{eq:USM-sHMF}.

\subsection{Wright-Fisher diffusion}
We now discuss the different families of methods that have been used to integrate Eq.~\eqref{eq:WFA}. 

The first class of methods is the usual algorithms for SDE, such as the Euler-Maruyama (EM) method or the Milstein method. This class of methods fails to capture the correct dynamics of Eq.~\eqref{eq:WFA} due to the non-Lipschitz multiplicative noise and the solution can leave the domain $[0,1]$ of Eq.~\eqref{eq:WFA}. 

The second class of methods introduces a min-max limiter 
\begin{equation}
\Theta(x) = \min(\max(x,0),1),
\end{equation}
 to project the numerical solution back onto $[0,1]$. The resulting methods are bounded and weakly convergent, but they are not strongly convergent. One can apply this limiter under the square root to get the \emph{internal limiter} (IL) method, see \cite{DoeringSargsyanSmereka05}, or to the complete update to get the \emph{external limiter} (EL) method.
 
 The third class of methods is based on the fact that Eq.~\eqref{eq:WFA} has an exact solution for particular values of the parameters. Moro and Schurz proposed a splitting method based on this idea, see \cite{MoroSchurz07}. The Moro-Schurz (MS) method has parameter restrictions, which limit its applicability.
 
 The fourth and last class of methods uses a control function to keep the solution in the bounded domain. This idea is due the Milstein \cite{MilsteinPlatenSchurz98} and can be used alone (\emph{balanced implicit method} (BIM), see \cite{MilsteinPlatenSchurz98}) or in conjunction with a splitting method (\emph{backward implicit split step} (BISS) method, see \cite{DangerfieldKayMacNamaraBurrage12}). These methods can be applied without restriction and can be shown to strongly converge. However, the rate at which the method converges is not known.

We now discuss the implementations of the different methods. Let us introduce $\Delta t$ a time increment and $\Delta W^n$ the $n$-th increment of a Wiener process. Then one can obtain the discrete approximation $x^n \approx x(t_n=n dt)$ of the different algorithms.

\begin{table}[t]
\caption{\label{tab:table1}This table summarizes the properties of the different numerical methods available to solve the Wright-Fisher diffusion equation. We consider whether the method produced a bounded result, is weakly  convergent or strongly convergent. In case of convergence, we specify whether there is restriction on parameters on not.}
\begin{ruledtabular}
\begin{tabular}{ccccc}
 Method&Bounded&Weak conv.&Strong conv.&No Restrict.\\ \hline\\[-0.6em]
 EM&$\times$&$\times$&$\times$&$\times$\\
 IL&$\times$&\checkmark&$\times$&$\times$\\
 EL&\checkmark&\checkmark&$\times$&$\times$\\
  MS&\checkmark&\checkmark&\checkmark&$\times$\\
 BIM&\checkmark&\checkmark&\checkmark&\checkmark\\
 BISS&\checkmark&\checkmark&\checkmark&\checkmark
\end{tabular}
\end{ruledtabular}
\end{table}

\begin{description}
\item[EM] The EM method is given by
\[
x^{n+1} = x^n -\gamma(x^n-\mu)\Delta t + \sigma \sqrt{x^n(1-x^n)}\Delta W^n.
\]
This method does not converge at all and leads to unrealistic results.
\item[Internal limiter (IL)] The IL method is defined as
\[
x^{n+1} = x^n -\gamma(x^n-\mu)\Delta t + \sigma \sqrt{\Theta(x^n)(1-\Theta(x^n))}\Delta W^n.
\]
This method is not bounded, but is weakly convergent.
\item[External limiter (EL)] The EL method is defined as
\[
x^{n+1} = \Theta\left(x^n -\gamma(x^n-\mu)\Delta t + \sigma \sqrt{x^n(1-x^n)}\Delta W^n\right).
\]
This method is bounded and weakly convergent, but not strongly convergent.
\item[Moro-Schurz (MS)] The MS method is based on the following splitting:
\begin{subequations}
\begin{eqnarray}
dy_1 &=&\frac{\sigma^2}{2}(y_1-\frac1{2})dt + \sigma \sqrt{y_1(1-y_1)}dW_t\\
dy_2 &=& \left[-\gamma(y_2-\mu)-\frac{\sigma^2}{2}(y_2-\frac1{2})\right]dt
\end{eqnarray}
\end{subequations}
The first equation has an exact solution. At each time step, the first equation is solved analytically and serves as an initial condition for the second equation, which is solved using a Forward Euler algorithm. This method is only bounded for certain parameters, for which it is both weakly and strongly convergent.

\item[BIM] The BIM is defined as
\[\begin{aligned}
x^{n+1} =& x^n -\gamma(x^n-\mu)\Delta t + \sigma \sqrt{x^n(1-x^n)}\Delta W^n\\ &+ D(x^n)(x^n-x^{n+1}),
\end{aligned}\]
where 
\begin{equation}
D(x^n) = d^0(x^n)\Delta t + d^1(x^n)|\Delta W_n|,
\end{equation}
 is a control function. The convergence of this method depends on the choice of $d^0$ and $d^1$.  For good control functions, this method is both weakly and strongly convergent. The limitation of this method is that it is not always clear how to choose the appropriate control functions.

\item[BISS] The BISS method is based on the splitting
\begin{subequations}
\label{eq:BISS}
\begin{eqnarray}
dy_1 &=& \sigma \sqrt{y_1(1-y_1)}dW_t\label{eq:BIM.1}\\
dy_2 &=& -\gamma(y_2-\mu)dt
\end{eqnarray}
\end{subequations}
and solves the first equation using the BIM and the second using an forward Euler step. In the BIM step, the function $d^0(x^n)\equiv 0$ and 
\begin{equation}\label{eq:control}
d^1(x) = \left\{
\begin{array}{ll}
\sigma \sqrt{\frac{1-\varepsilon}{\varepsilon}}&\text{if } y<\varepsilon,\\
\sigma \sqrt{\frac{1-y}{y}}&\text{if } \varepsilon \leq y<\frac12,\\
\sigma \sqrt{\frac{y}{1-y}}&\text{if } \frac12<y\leq1-\varepsilon,\\
\sigma \sqrt{\frac{1-\varepsilon}{\varepsilon}}&\text{if } y>1-\varepsilon,
\end{array}
\right.
\end{equation}
where $\varepsilon$ is a small tolerance parameter, defined in \cite{DangerfieldKayMacNamaraBurrage12} as
\begin{equation*}
\varepsilon = \min(A\Delta t,B\Delta t,1-A\Delta t,1-B\Delta t) >0, 
\end{equation*}
for $\Delta t$ small enough and where $A=\gamma\mu$ and $B=\gamma(1-\mu)$.
The discretization of Eq.~\eqref{eq:BISS} takes the form
\begin{subequations}
\begin{eqnarray}
y_*^{n+1} &=& y_1^{n} + \frac{\sigma\sqrt{y_1^{n}(1-y_1^{n})}\Delta W_n}{1+d^1(y_1^n)|\Delta W_n|},\\
y_1^{n+1} &=& y_*^{n+1} -\gamma(y_*^{n+1}-\mu)\Delta t.
\end{eqnarray}
\end{subequations}
 This method is bounded and converges weakly and strongly for all parameters if $\Delta t$ is chosen small enough.

\end{description}

The characteristics of the different methods are summarized in Tab.~\ref{tab:table1}. The two best methods are the BIM and the BISS. We choose the BISS because of it is easier to adapt to more complex dynamics such as the dynamics of the sHMF. The BIM could also be used, the problem is that for more complex dynamics, a good control function $d^0$ is difficult to define. Since the convergence rate of the BISS is unknown, we expect numerical artifacts close to the boundaries of the domain, where the Lipschitz condition is not satisfied.

\subsection{Numerical methods for the sHMF of the USM}
In the first part of this appendix, we have recalled the numerical methods available for solving the WF diffusion equation. For the sHMF of the USM Eq.~\eqref{eq:USM-sHMF}, one needs to deal with the noises of all neighbouring degree classes. This can be done by a splitting method inspired by the BISS algorithm. We describe it for two variants $V=2$. The idea is to split the update between the utterance production (which is noisy) and the deterministic learning rule. The continuous time version of the normal approximation Eq.~\eqref{eq:HMFutterance} is obtained by scaling $\frac1{E}= dt$. The first component $u_1$ is of the form
\begin{equation}\label{eq:NA-cont}
u_1 = a + b\left[x_1 + \sigma_k\sqrt{x_1(1-x_1)}\frac{\Delta W_n}{\Delta t}\right],
\end{equation}
where $\sigma_k = (kLN_k)^{-1/2}$, $a=m_1$ and $b=1-m_1-m_2$ for a matrix $M$ defined by Eq.~\eqref{eq:M}.

The idea is to modify Eq.~\eqref{eq:NA-cont} by introducing the control function $d^1$ of Eq.~\eqref{eq:control}, leading to the utterance production
\begin{subequations}\label{eq:sHMF-BISS}
\begin{equation}
u^{n+1}_1 = a + b\left[x^n_1 + \frac{\sigma_k\sqrt{x^n_1(1-x^n_1)}\frac{\Delta W_n}{\Delta t}}{1+d^1(x_1)|\Delta W_n|}\right],
\end{equation}
and the learning update given by Eq.~\eqref{eq:HMFdx}
\begin{equation}\begin{aligned}
x_1^{(k),n+1} =\ & x_1^{(k),n}+ \lambda(1-h)k(u_1^{(k),n+1}-x_1^{(k),n})\Delta t \\&+ \lambda hk\sum_{k'}p(k'|k)( u_1^{(k'),n+1}-x_1^{(k),n})\Delta t.
\end{aligned}\end{equation}
\end{subequations}
Eq.~\eqref{eq:sHMF-BISS} is the BISS algorithm for the sHMF approximation of the USM. This approximation ensures that $x_1\in [0,1]$, for all degree classes. The strong convergence remains to be shown, but since the BISS is strongly convergent, we have good reason to think that this algorithm conserves this property. For $V>2$, the same idea can be used. The only difficulty is to find an appropriate control function.

\bibliography{Project}

\end{document}